# Generation of Isocurvature from Curvature Inhomogeneities on Super-Horizon Scales

**Albert Stebbins**
Theoretical Astrophysics Department
MS209
Fermi National Accelerator Laboratory
Box 500
Batavia, IL 60510
USA
stebbins@fnal.gov

**Abstract**

*Here it is shown 1) how isocurvature inhomogeneities correlated on causally disconnected (super-horizon) scales are generated from curvature inhomogeneities which are known to be correlated on these scales 2) that super-horizon isocurvature generation is nearly inevitable for non-equilibrium chemical processes 3) that the amplitude of the compositional isocurvature correlations a) can be large for production of rare objects, b) falls off rapidly with separation b) falls off at scales below the horizon when these modes are generated. These two fall-offs results in an "isocurvature bump" in the power spectrum. Isocurvature generation is illustrated by the process of dark matter freeze-in, computed here with both separate universe modelling and linear perturbation theory. For freeze-in the most prominent isocurvature modes are inhomogeneities in the ratio of dark matter to standard model matter. Much smaller inhomogeneities in the ratio of baryons to standard model entropy are also produced. Previous constraints on freeze-in from Ly-α clouds limit the bump enhancement to ≲ 10% on comoving scales ≲ 1 Mpc. Current observations are not sensitive to the isocurvature modes generated in viable freeze-in models. Results are obtained using a somewhat novel framework to describe cosmological inhomogeneities.*

November 29, 2023

# I. Introduction

This paper elucidates the physical mechanisms by which cosmological curvature inhomogeneities can evolve into a mixture of curvature and isocurvature inhomogeneities on scales much larger than the cosmic horizon. It is argued that such evolution is inevitable for chemical processes (processes that change the composition of the matter) to the extent that these processes are out of equilibrium (deviations from detailed balance). For example dark matter freeze-in or freeze-out, big bang nucleosynthesis or production of primordial black holes. Such late-time isocurvature remnants of processes which occur in the early universe would provide novel information about these processes. This study was occasioned by an early claim by Bellomo Berghaus & Boddy (2022 arXiv:2210.15691v1 hereafter BBB22), that in some cases, dark matter produced by a "freeze-in" mechanism would result in the generation of super-horizon isocurvature inhomogeneities seeded by preexisting curvature inhomogeneities; and at a level greater than allowed by observations of the angular spectrum of CMB anisotropies. This controversial claim has been since been superseded by the published Bellomo Berghaus & Boddy (2023). This paper resurrects some of but not all of the initial claims. Spirited discussions surrounding these issues have led this author to feel there are common misunderstandings surrounding a number of topics in cosmology which has motivated a large pedagogical section (§II) to this paper. To paraphrase an argument that has been made a few times: "Since slow role inflation has a single clock there is no correlated information on super-horizon scales; correlated inhomogeneities only arise at horizon crossing". It is explained below how this erroneous. While such misunderstandings can lead to erroneous conclusions about isocurvature generation any common misunderstanding should be cleared up for its own sake. A novel approach to these topics is described here and it is hoped that some reader find this approach more intuitive while at the same time being more rigorous than the standard approach.

**Ia The Short Version**

A (controversial) claim was and is that isocurvature modes, e.g. inhomogeneities in the ratio of the density of dark matter relative to the density of photons, can be produced on super-horizon scales and are proportional to the curvature inhomogeneities. This is indeed true. For curvature inhomogeneities to seed isocurvature inhomogeneities on super-horizon scales requires a mechanism whereby physical process could result in differing yields of dark matter in causally disconnected parts of the universe. While freeze-in "events" in widely separated regions of the universe are indeed causally disconnected they also share correlated information. To be explicit about this information consider two separated streamlines of matter in an expanding universe after inflation. Restrict consideration to these streamlines in the time interval from the end of inflation to the times where the past light cones of the two streamlines intersect at the end of inflation. In this interval one can say that the two streamlines are causally disconnected, that nothing that happens in one segment can have causal effect in the other. In each of these two streamline segments one can identify an event where the matter density takes some particular value, $\rho$. At these two events determine the rate of expansion, $\theta_1$ and $\theta_2$, of the cosmological flow (for spherical expansion $\theta = 3H$ where $H$ is the Hubble expansion rate). These two expansions should be close to the Einstein-deSitter (flat) universe value, $\theta_{\text{EdS}} \equiv \sqrt{24\pi G \rho}$. However since there are curvature inhomogeneities there will be fluctuations about $\theta_{\text{EdS}}$: $\theta_1 \neq \theta_{\text{EdS}}$ and $\theta_2 \neq \theta_{\text{EdS}}$. Our universe is observed to be nearly flat with a (nearly scale invariant / Harrison-Zel'dovich) spectrum of curvature inhomogeneities such that the curvature is correlated on super-horizon scales. This means that causally disconnected $\theta_1$ and $\theta_2$ will be correlated, i.e. $\langle \theta_1 \theta_2 \rangle - \langle \theta_1 \rangle \langle \theta_2 \rangle \neq 0$. This non-zero covariance for causally disconnected points is a super-horizon covariance which can have super-horizon manifestations.

The mechanism by which super-horizon covariances from curvature inhomogeneities lead to super-horizon covariances of compositional isocurvature inhomogeneities can be explained fairly simply. Non-thermal cosmological "reactions" are interactions between particles that while evolving toward thermal equilibrium do not run to completion resulting in a non-thermal final state. The approach to equilibrium is frustrated by the finite cosmological time: the universe expands diluting particle density and lowering temperature both of which decrease the effectiveness of reactions which act to keep the fluid in equilibrium. Eventually these reactions become inactive leading to a frozen final state usually reflected in the relative abundance of various particle species. The local curvature modulates the final frozen state through it's effect on the local rate of expansion of the cosmological fluid. At a given temperature/density a region of greater curvature has slightly slower expansion than a region of lower curvature causing the cosmological fluid to spend more proper time at a given temperature/density, allowing the reactions to proceed further toward equilibrium. For dark matter freeze-in the dark matter is initially under-abundant so reactions increase the dark matter abundance towards thermal equilibrium. Thus dark matter abundance is positively correlated with curvature for freeze-in. The opposite is true for dark matter freeze-out where dark matter becomes over abundant as the universe cools so annihilations act to decrease its abundance. This decrease will be enhanced by positive curvature resulting in a dark matter abundance anti-correlated with curvature. For the case of big bang nucleosynthesis where light elements become over-abundant relative to heavy elements regions of larger curvature are able to produce more heavy elements, not only leading to small chemical abundance inhomogeneities but also smaller baryon mass density since heavy elements are more bound and carry very slightly less mass per baryon than lighter elements. The super-horizon nature of these inhomogeneities is the statement that the final yield, $Y_f$, e.g. ratio of dark matter particles to baryons has non-zero covariance at $\langle Y_f^1 Y_f^2 \rangle - \langle Y_f^1 \rangle \langle Y_f^2 \rangle \neq 0$ for causally disconnected cosmic fluid elements. All of these effects are extremely small; in part because the during early times where these reactions take place the fractional difference in the expansion rate between high and low curvature regions is extremely small. In this sense a statement that no isocurvature inhomogeneities are produced is not very wrong. The isocurvature amplitude depends on how sensitive the reaction rate is to the expansion rate. One can imagine cosmic processes where this dependence is much stronger than for the processes described above yielding much larger isocurvature inhomogeneities.

### Ib Outline

Much of this paper is of a pedagogical nature (true but not new). The reader can skip the pedagogy in §II and go to §III which gives general semi-quantitative arguments about the generation of isocurvature inhomogeneities, and §IV which quantitatively derives the isocurvature production by dark matter "freeze-in". In §IVa this is first formulated in the "separate universe" formalism and in §IVb in terms of perturbative inhomogeneities. §IVc analyzes the results of §IVa&b to obtain simple and useful expressions for the isocurvature produced by freeze-in. §IVd focuses on the baryon density isocurvature modes. §IV contains a semi-quantitative discussion of observational constraints on freeze-in. §V gives a summary of key points. Appendix A gives analytic approximations to the reaction rates for millicharged dark matter freeze-in as derived BBB22.

### Ib Notation, Conventions and ΛCDM Cosmology

HEP natural units $c = \hbar = k_B = 1$ are used except when they're not. Conventional tensor index notation with semicolon covariant derivatives and metric signature $\{-, +, +, +\}$ are used. The matter is divided into two types, standard matter, abbreviated SM, and dark matter abbreviated DM means. DM refers to the entire dark sector. Dark matter particle refers to the particle which constitutes the dark matter which has mass $m_d$. Baryons are a constituent of SM and baryon number refers to the baryon asymmetry not the total absolute number of baryons. We use a prescript "s" for SM, "d" for DM and "b" for baryon number. To avoid notational difficulties with tensor indices these scripts proceeds the symbol for scalar fields and tensors but follow for constants and functions. Predictions will be set in the context of a concordance ΛCDM cosmology as described in §IVc3.

# II. Pedagogy

## IIa FLRW

The cosmic reactions under consideration occurs in the context of small perturbations from exact cosmological spacetimes which are here referred to as FLRW in recognition of their originators. Next we review some basics of FLRW cosmology which are easily generalized to these perturbed spacetimes. The key properties of these cosmologies is the cosmological principle of homogeneity and isotropy which is exact for FRLW. FLRW spacetimes have metric given by

$$ds^2 = -dt^2 + a[t]^2 \gamma_{ij} dx_i dx_j \quad (1)$$

where $a[t]$ is the scale factor defined up to a multiplicative constant and $t$ is the proper time in the cosmic frame which is defined up to an additive constant. The isometric (homogeneous and isotropic) 3D hypersurfaces have metric $a[t]^2 \gamma_{ij}$ where $\gamma_{ij}$ is the metric of $\mathbb{S}^3, \mathbb{E}^3, \mathbb{H}^3$ for $\text{sign}[k] = +1, 0, -1$, respectively. In general relativity (GR) the matter content of FLRW spacetimes is always a perfect fluid with homogeneous mass density $\rho[t]$ and isotropic pressure $p[t]$. The scale factor evolves according to the Friedmann equation

$$\left(\frac{\dot{a}[t]}{a[t]}\right)^2 = \frac{8\pi G \rho[t]}{3} - \frac{k}{a[t]^2} \qquad \dot{\rho}[t] = -3 \frac{\dot{a}[t]}{a[t]} (\rho[t] + p[t]) \quad (2)$$

where the latter ensures local energy conservation. The over dots indicate derivatives wrt $t$ which is the proper time in the cosmic frame Any cosmological constant has been absorbed into $\rho$ and $p$.

## IIb Kurvature

### IIb1 Specific Binding Energy

The $\gamma_{ij}$ have intrinsic curvature $\text{sign}[k]$ so the FLRW isometric hypersurfaces have a time dependent intrinsic curvature $K[t] = \frac{k}{a[t]^2}$. The quantity $K[t]$ is not only the geometrical curvature of a 3-surface but also a measure of gravitational binding energy. As illustrated the formula $K[t] = \frac{8\pi G \rho[t]}{3} - \left(\frac{\dot{a}[t]}{a[t]}\right)^2$ it is the difference of a term related to the gravitational energy and a term related to the kinetic energy. More precisely in a uniformly expanding uniform medium a Newtonian measure of how bound two particles separated by a physical distance $r$ is the specific binding energy $b \equiv \frac{GM}{r} - \frac{1}{2} v_{rel}^2$ where, $v_{rel}$ is their relative velocity and $M$ is the mass within a sphere of radius of $r$. In FLRW spacetimes $r[t] = a[t] r_{co}$, $v_{rel} = \frac{\dot{a}[t]}{a[t]} r[t] = \dot{a}[t] r_{co}$ and $M[t] = \frac{4\pi}{3} \rho[t] r[t]^3$. Defining the comoving separation $r_{co} = \frac{r[t]}{a[t]}$ which is constant in time one finds $b = \frac{1}{2} K[t] a[t]^2 r_{co}^2 = \frac{1}{2} k r_{co}^2$. Positive (negative) $b$, or $K$ or $k$ indicate positive (negative) binding. The values of $k$ and $r_{co}$ depend on the arbitrary normalization of $a[t]$ and though $k r_{co}^2$ does not depend on this normalization it does depend on the arbitrary choice of particles. A good way to quantify this binding is $K$ which is measurable since it depends only on $\rho$ and $\frac{\dot{a}}{a}$ both of which are covariant. Below $K$ is generalized to a nearly arbitrary spacetimes. $K$ is related to the density parameter $\Omega \equiv \frac{8\pi G \rho}{3}\left(\frac{a}{\dot{a}}\right)^2$ by $K = \left(\frac{\dot{a}}{a}\right)^2 (\Omega - 1)$. An advantage of using $K$ rather than $\Omega$ is it's simple relation to the scale factor $K[t] \propto a[t]^{-2}$ and through this to various conserved covariant quantities such as particle number density or entropy density $K[t] \propto n[t]^{2/3} \propto s[t]^{2/3}$. Referring to $K$ as curvature misses it's main utility which is its measure of gravitational binding. Also the word "curvature" is not very descriptive as very many things are curvature in metric theories of gravity. Furthermore "curvature" also has different meanings in among cosmologists. For this reason we will henceforth refer to $K$ and its generalizations as *kurvature*.

## IIb2 Kurvature of the Space-Time Fluid

Since $K$ plays such an important role in this paper and is extracted from the specific FLRW class of space-times we now show how it may generalized to a local covariant quantity in nearly any space-time. Using Einstein's equation one may make a fluid decomposition of the Ricci curvature tensor:

$$R^{\alpha\beta} = G^{\alpha\beta} - \frac{1}{2} g^{\alpha\beta} G^{\gamma}{}_{\gamma} \qquad G^{\alpha\beta} = 8\pi G\, T^{\alpha\beta} \qquad T^{\alpha\beta} = (\rho + p)\, u^{\alpha} u^{\beta} + p\, g^{\alpha\beta} + q^{\alpha\beta} \tag{3}$$

where $u^{\alpha}$ is the center-of-momentum 4-velocity while $\rho, p$ and $q^{\alpha\beta}$ are the density pressure and anisotropic stress in the center-of-momentum frame ($u^{\alpha} u_{\alpha} = -1$, $q^{\alpha\beta} u_{\beta} = 0$, $q^{\alpha\beta} = q^{\beta\alpha}$, $q^{\alpha}{}_{\alpha} = 0$). This decomposition is determined by the unique solving the eigenvalue $T^{\alpha}{}_{\beta} u^{\beta} = -\rho\, u^{\alpha}$. Apart from a choice of time forward direction this decomposition is unique on time orientable space-times so long as $T^{\alpha\beta}$ has no Lorentz boos symmetries. If $T^{\alpha\beta}$ has Lorentz symmetries, e.g. vacuum or deSitter space, the kurvature concept ceases to be useful. This fluid representation of the Ricci curvature is here referred to as the *space-time fluid* which is the *aggregate* of all matter. The space-time fluid has a rate of expansion, $\theta \equiv u^{\alpha}{}_{;\alpha}$, and with this one has a generalized definition of kurvature:

$$K \equiv \frac{8\pi G \rho}{3} - \frac{1}{9}\theta^2. \tag{4}$$

In FLRW space-times, since $\theta = 3\frac{\dot{a}}{a}$, it reduces to the usual formula. From this definition the evolution of $K$ can be derived, using among other identities the Raychaudhuri equation (Hawking & Ellis 1973) and is given by

$$\dot{K} = -\frac{2}{3}\theta K - \frac{2}{9}\theta\left(\mathcal{P}^{\alpha\beta}\dot{u}_{\alpha;\beta} + \dot{u}_{\beta}\dot{u}^{\beta}\right) - \frac{8\pi G}{3}\sigma_{\alpha\beta}q^{\alpha\beta} \qquad \dot{u}^{\alpha} = -\frac{\mathcal{P}^{\alpha\beta}(p_{;\beta}+q_{\beta\gamma}{}^{;\gamma})}{\rho+p} \tag{5}$$

where $\dot{K} \equiv u^{\alpha} K_{;\alpha}$ and $\dot{u}^{\beta} \equiv u^{\alpha} u^{\beta}{}_{;\alpha}$ respectively give the convective proper time derivative of $K$ and $u^{\alpha}$ along the $u^{\alpha}$ flow lines. The vector $\dot{u}^{\beta}$ is the proper acceleration. The idempotent (projection) tensor $\mathcal{P}^{\alpha\beta} \equiv u^{\alpha} u^{\beta} + g^{\alpha\beta}$ removes the components of tensors which are temporal in the fluid frame, $u^{\alpha}$. The tensor $\sigma_{\alpha\beta} \equiv \frac{1}{2}\mathcal{P}_{\alpha}{}^{\gamma}\left(u_{\gamma;\delta} + u_{\delta;\gamma} - \frac{1}{2}g_{\delta\gamma}u^{\eta}{}_{;\eta}\right)\mathcal{P}^{\delta}{}_{\beta}$ is the rate of shear tensor for the space-time fluid (Hawking & Ellis 1973).

## IIb3 Kurvature and 1st Order Cosmological Perturbations

Using the convention that spatial pressure gradients are neglected in the super-horizon approximation one can demonstrate the conservation of the curvature potential, $\mathcal{R}$, on super-horizon scales. Define $O[n, m]$ to represent any expression where $n$ gives the lowest order of perturbation from cosmological homogeneity and isotropy in the expression and $m$ gives the smallest number of spatial pressure gradients in an expression. Use $\dot{u}_{\alpha\beta} \sim q_{\alpha\beta} \sim \sigma_{\alpha\beta} \sim O[1, 0]$ since by symmetry (homogeneity and isotropy) these quantities are zero in FLRW spacetimes. Thus $\theta = 3\frac{\dot{a}}{a} + O[1, 0]$, $\mathcal{P}^{\alpha\beta} p_{;\beta} = O[0, 1]$, $\mathcal{P}^{\gamma\delta} q_{\beta\gamma;\delta} = \mathcal{P}^{\gamma\delta} q_{\beta\gamma;\delta} = O[1, 1]$, $\dot{u}^{\alpha} = O[0, 1]$ $\dot{u}_{\beta} \dot{u}^{\beta} = O[0, 2]$, $\mathcal{P}^{\alpha\beta} \dot{u}_{\alpha;\beta} = O[0, 2]$ and finally $\dot{K} = -2\frac{\dot{a}}{a} K + O[1, 2]$. Hence $K \propto a[t]^{-2} + O[1, 2]$. Note that the somewhat fuzzily defined specific binding energy $b$ introduced above is $\propto a^2 K$. Thus heuristically one finds that the specific binding energy $b$ is constant on super-horizon scales to 1st order in cosmological perturbation theory. More specifically in 1st order gauge-invariant perturbation theory of flat ($k = 0$) one has (Weinberg 2008)

$$K = \frac{2}{3}\frac{1}{a^2}\nabla_{co}^2 \mathcal{R} + O[2] \qquad \mathcal{R} \equiv \Psi + \frac{2}{3}\frac{\bar{\rho}}{\bar{\rho}+\bar{p}}\left(\Phi + \frac{a}{\dot{a}}\frac{\partial}{\partial t}\Psi\right) \tag{6}$$

where $\nabla_{co}^2 = \frac{\partial}{\partial \boldsymbol{x}} \cdot \frac{\partial}{\partial \boldsymbol{x}}$ is the Laplacian wrt the comoving spatial coordinate, $\boldsymbol{x}$, $\mathcal{R}$ is the curvature potential, $\Phi$ and $\Psi$ are the Bardeen potentials (Bardeen 1980), while $\bar{\rho}$ and $\bar{p}$ are the "background" density and pressure. Thus since $K \propto \mathcal{R}/a^2$ we see that $\mathcal{R}$ is constant on super-horizon scales, or more specifically the Fourier amplitude $\tilde{\mathcal{R}}[\eta, \boldsymbol{k}]$ is approximately constant when $|\boldsymbol{k}|\eta \ll 1$ where $\eta$ is the conformal time. $K$ is related to but not the same as $\mathcal{K} \equiv 6K + \sigma^2$ used in Ellis & Bruni (1989) where $\sigma^2 \equiv \frac{1}{2}\sigma_{\alpha\beta}\sigma^{\alpha\beta}$ (N.B. definitions of $\sigma$ vary by numerical factors in different texts).

While $K$ is a local covariant quantity $\Phi$, $\Psi$ and $\mathcal{R}$ are not - though they are "gauge invariant" in the sense they are invariant to small changes in coordinates. $\Phi$, $\Psi$ and $\mathcal{R}$ are dimensionless while $K$ has dimensions of inverse length squared. One can construct various local covariant dimensionless quantities which are determined by $K$:

$$\Omega \equiv \frac{24\pi G \rho}{\theta^2} = 1 + \frac{9K}{\theta^2} \qquad \kappa \equiv \frac{3K}{8\pi G \rho} = \frac{\Omega - 1}{\Omega} \tag{7}$$

The former is the well-known local density parameter. The latter, which has no commonly used name, is a useful small dimensionless parameter.

## IIc Barotropy and Adiabaticity

In freeze-in models one typically divides the matter into two components, here labeled by SM (standard matter) and DM (dark matter), which are here treated as fluids with a density $\rho$ and isotropic pressure $p$. The SM component, which might or might not correspond to Standard Model particles, is strongly interacting and remains very close to thermal equilibrium. This thermal fluid does have a non-zero chemical potential giving the net baryon asymmetry. One can parameterize the SM EoS as $p_s[_s\rho, _b\eta]$ where $_b\eta$ (defined below) parametrizes the baryon asymmetry. It is assumed that $_b\eta$ is initially uniform throughout the fluid and to the extent that it remains that way the SM is *barotropic*, meaning pressure depends only on density and is therefore well described by a universal EoS. Barotropic fluids are common in cosmology usually as a manifestation of cosmological homogeneity.

Energy momentum is always locally conserved. Usually on super-horizon scales in cosmology where all fluids are comoving, one can treat the aggregate of all the fluids as single (space-time) fluid (eq. 3) which together locally conserves energy momentum. On super-horizon scales gradients of temperature and velocity are small so processes such as thermal conduction or viscosity which cause anisotropic pressure are neglected and thus the aggregate fluid element is close to a perfect fluid which does not transfer energy-momentum between fluid elements only gaining or loosing energy through expansion or contraction via $p\,dV$ work. Such a fluid is called *adiabatic*. The aggregate fluid is always adiabatic on super-horizon scales and often in cosmology the individual fluids which make up the aggregate are themselves adiabatic. Adiabaticity does not imply barotropy and barotropy does not imply adiabaticity.

For matter models considered here there is energy exchange between SM and DM so neither fluid is adiabatic but, as will be shown below, non-adiabaticity of the SM fluid may often be neglected. For these matter models it is assumed that the aggregate fluid is initially barotropic. No conservation law prevents the aggregate fluid from becoming non-barotropic even on super-horizon scales i.e. evolving to different EoSs in separate horizon patches, however there must be some mechanism which leads to inhomogeneity in the aggregate EoS. If the energy exchange from SM to DM does not preserve $_b\eta$ then the fluid EoS will change $p_s[_s\rho, _b\eta]$ but if $_b\eta$ depends only on $_s\rho$ the SM is still barotropic. Below it is shown how super-horizon kurvature inhomogeneities causes inhomogeneities in the SM to DM energy exchange which results in inhomogeneities in $_d\rho/_s\rho$ and in $_b\eta$. Super-horizon inhomogeneities in $_b\eta$ means the SM will become non-barotropic on super-horizon scales which may be described as "baryon-to-photon" inhomogeneities. Furthermore since $_dp/_d\rho \neq _sp/_s\rho$ inhomogeneities in $_d\rho/_s\rho$ mean that the aggregate fluid pressure $_sp + _dp$ will not only depend on the aggregate fluid density $_s\rho + _d\rho$, i.e. there is no aggregate EoS. In this sense the aggregate SM+DM fluid (space-time fluid) will also be abarotropic on super-horizon scales. This may be described as "dark matter-to-photon" inhomogeneities. Both of these are observable in the measured CMB temperature and polarization anisotropies. More abstractly the phenomena described in this paper is the seeding of super-horizon abarotropy (non-barotropy) by kurvature inhomogeneity.

Cosmologists almost always use (or misuse) the term "adiabatic" or "isentropic" to describe what is probably better described as barotropic; and conversely "non-adiabatic" or "anisentropic" to describe abarotropy. The idea behind the use of "adiabatic" is that the matter inhomogeneities are "the same as if" one takes the background cosmology, assumed to consist of a set of comoving spatially uniform barotropic perfect fluids, $_x\bar\rho$ and $_x\bar p = p_x[_x\bar\rho]$, and at each point compresses or expands each of the fluids adiabatically. In this case, for small inhomogeneities, $_x\delta\rho \cong -3\,(_x\bar\rho + p_x[_x\bar\rho])\,\delta\ln[V]$ where $\delta\ln[V]$ is the change in log volume, and thus $_x\delta\rho/(\bar\rho_x + p_x[\bar\rho_x])$ is the same for all fluids. Due to the assumed barotropy of each fluid $_x\delta p \cong _xc_s^2\,_x\delta\rho$ where $_xc_s^2 \cong p_x'[_x\bar\rho] = _x\dot{\bar p}\big/_x\dot{\bar\rho}$ is the adiabatic sound speed. Another consequence of this adiabatic squeezing is that the aggregate fluid is barotropic: $\delta\bar p/\delta\bar\rho = \dot{\bar p}\big/\dot{\bar\rho}$. In other words the consequence of this "adiabatic" assumption is barotropy. One then might argue that either adiabatic or barotropic are good descriptors of these inhomogeneities. The problem with this usage is twofold: 1) "the same as if" descriptions are not necessarily what is actually happening, 2) if one uses the word "adiabatic" to describe what might have happened then one no longer has this word available to describe what is happening, i.e. whether or not energy is transported into or out of fluid elements. This "dynamical adiabaticity" is how the word is used in fluid mechanics. A fluid may be dynamically adiabatic whether or not it is barotropic and it is problematic to not have this word to describe this. It is better to describe "what is" (barotropy) than to describe "what might have been" (adiabaticity). It could be that the fluids are barotropic because they behaved adiabatically in which case the adiabatic descriptor is appropriate however the term is used more generally. One finds that "adiabatic" is shorthand for the mathematical assumption that the total entropy perturbation, usually denoted $S$, is zero. However the condition for $S = 0$ is that the aggregate fluid is barotropic and the anisotropic stress is negligible; the "why" of barotropy and isotropy doesn't matter. Fluids may be barotropic for other reasons e.g. they're ultra-relativistic so $p = \rho/3$ or they are constrained to a fixed equation of state by, say, thermal equilibrium. This latter reason is the approximation for the SM fluids used below. Since total energy is always locally conserved the aggregate fluid is always dynamically adiabatic. Saying "the cosmological fluid is adiabatic" is tautological in the sense that the aggregate fluid always exhibits dynamical adiabaticity. Such statements are a source of confusion because they contains no useful information . Yet such statements such as there are common in the literature.

The dictionary definition of isentropic is "constant entropy". This might be a better descriptor for what is commonly called adiabatic in cosmology *except* for the fact that "isentropic flow" generally refers to processes which are both adiabatic and reversible. Since in cosmology adiabaticity almost always implies reversibility isentropic and adiabatic flows are effectively synonymous. Thus isentropic is not an ideal alternative. "Barotropic" or "pure curvature" inhomogeneities are better descriptors of what is commonly called "adiabatic". "Pure curvature" is sometimes used and gives a clearer dichotomy of different types of inhomogeneity: curvature and isocurvature. Nevertheless the use/misuse of the word "adiabatic" pervades the literature and is used in almost all textbook descriptions of cosmological inhomogeneities

## IId Local Covariant Clocks

In the theory of cosmological inflation there is the concept of "single clock inflation" which is essentially equivalent to the slow roll approximation where there is a fixed relation between the inflation field and it's time derivative. The single clock is meant to indicate that there is only a single parameter describing the inflaton field which governs the growth of curvature perturbations. Here a *different* definition of "clock" is used: a clock is any time varying quantity associated with a space-time point which one can use to identify the time, either of the universe as a whole for FLRW space-times or locally in an inhomogeneous space-time. Clocks which are physically measurable quantities localized in space and time are called *local covariant* clocks. These are usually scalars in the tensor sense. Two clocks with fixed functional relationship with no free parameters are *synchronized* clocks. Clocks which are not synchronized are *independent*. Quantities in equations often used to describe the cosmic evolution are often not measurable and even when they are may not be local. Examples of non-covariant clocks are 1) the conformal or other sorts of non-covariant coordinate times 2) the scale factor, $a$, which is only defined up to a multiplicative factor, 3) the proper time which is only defined up to an additive offset. The density, pressure, entropy density and temperature of a barotropic gas are all local covariant clocks but for barotropic fluids they are all synchronized and not independent.

To describe physical (chemical) processes such as freeze-in conversion of SM to DM it is useful to change the time variable from a non-covariant clock (say coordinate $t$) to covariant clock, (say $\rho$) so that the rate equations are themselves completely covariant. It is interesting to count the number of independent covariant clocks. If there is only one such clock then the yield of the chemical process will depend only on this clock value and if chemical process proceeds only for a finite range of clock values then the final yield in traversing that range will be a fixed number, the same everywhere, and thus homogeneous. This argument will be made more specific below but as an example: if dark matter freeze-in rates depends only on temperature and proceed only for a finite range of temperatures then if temperature is the only clock and every fluid element evolves from a high temperature above that range to a low temperature below that range then the dark matter yield will be the same for every fluid element. This argument is used by Weinberg (2008) to show that if all the matter in the universe completely thermalizes to a fixed EoS then then isocurvature inhomogeneities are erased. This paper deals with the case where there is not complete thermalization. In cosmologies there are cuvature inhomogeneities so there are at least two independent covariant clocks, $\rho$ and $\theta$. These two independent clocks allow for spatial differences in the $\rho$ vs. $\theta$ relationship to cause spatially inhomogeneous yields develop from non-thermal processes.

Roughly speaking $\rho$ gives the temperature and $\theta$ gives the rate of change of temperature so varying $K$ will modify the temperature time relation allowing dark matter production to produce more or less dark matter particles per photon in some regions than others. For chemical reactions which do not proceed all the way to thermal equilibrium the temperature-time relationship is fundamental in determining the final state: the greater the time at a fixed temperature the closer one will get to an equilibrium state. A similar argument can be applied for systems going out of equilibrium such as dark matter freeze-out: the greater the time at a given temperature the less out-of-equilibrium the system will get. As will be shown the lower the redshift the greater are the inhomogeneities in the temperature time relationship and the more inhomogeneous will be the final state of out-of-equilibrium. processes.

## IIe Isocurvature Inhomogeneities

The simplest and apparently a reasonably accurate description of our inhomogeneous universe at early times or on super-horizon scales is that it's content is described by one or more comoving adiabatic barotropic vorticity-free nearly isotropically expanding perfect fluids in a near FLRW spacetime i.e. pure curvature inhomogeneities. Comoving means any initial relative velocity between fluids is negligible. Such a matter model has only a single way for it to be inhomogeneous. This type of curvature inhomogeneity corresponds to inhomogeneities in $K$. When significant deviations from this "curvature only" paradigm exist (other than non-zero vorticity) there is said to be "isocurvature fluctuations" since in addition to the curvature inhomogeneities there are additional inhomogeneities which leave the curvature homogeneous. Curvature and isocurvature fluctuations are scalar inhomogeneities while vorticity is a vector inhomogeneity and is not discussed here.

Isocurvature is not a very descriptive term since there are a variety of types of scalar inhomogeneities which deviate from the curvature paradigm (Bucher, Moodley & Turok 2000). Within the context of perfect fluids one can characterize these inhomogeneities as
- abarotropic: any or all of the fluids may be abarotropic

- compositional: the density ratios of any or all pairs of fluids is inhomogeneous
- velocity: any or all pairs of fluid are not comoving

For a single fluid only abarotropic isocurvature is possible. If all the fluids are comoving then, when one may combines the different fluid into a single aggregate, this aggregate can only have abarotropic isocurvature inhomogeneities in addition to curvature inhomogeneities. Even if all the fluids are barotropic the aggregate will be abarotropic if there are compositional inhomogeneities and $p/\rho$ of different fluids differ. Such an initial state might bec considered generic considering Weinberg's argument that thermalization erases isocurvature inhomogeneities which might have existed previously. However thermalization does not prevent isocurvature inhomogeneities from being generated after thermalization.

One should also make a distinction between isocurvature modes which have an evident gravitational manifestation and those that do not. In GR gravity is given by space-time curvature which is associated with the matter through the stress-energy tensor, $T^{\alpha\beta}$, via Einstein's equation. Only the total (aggregate) stress-energy gravitates. Abarotropic aggregates of perfect fluids also gravitate through different pressure density relations which contribute to $T^{\alpha\beta}$. This gravitational manifestation can be solely described as an entropy perturbation $S$. Compositional isocurvature may contribute little or not at all to $S$. This does not mean it doesn't exist or does not lead to observable effects but merely that it doesn't gravitate. For example baryons and dark matter are nearly gravitationally indistinguishable when the different coupling of baryons to photons is unimportant. However baryon density inhomogeneities are accompanied by equal electron inhomogeneities and electron inhomogeneities can be detected via the consequent inhomogeneities in the rate of Thomson scattering of photons by electrons. Gravitational inhomogeneities are not always a good measure of isocurvature inhomogeneities which could be observable.

## IId Entropy

For a perfect fluid with an EoS it is useful to define an "entropy" density (in the fluid frame) which for the SM is

$$s[\rho, \ldots] \equiv \mathrm{Exp}\left[\int^p \frac{d\rho}{\rho + p[\rho, \ldots]}\right] \ . \tag{8}$$

There is multiplicative constant or "normalization" in this definition which is initially left unspecified. This definition of entropy is not necessarily related to thermodynamic entropy and even if it does one does not have to use the thermodynamic normalization. Entropy density is a useful quantity since for adiabatic fluids entropy is conserved, $(s\, u^\alpha)_{;\alpha} = 0$ or $\dot{s} \equiv u^\alpha s_{;\alpha} = -s\, \theta$. Thus entropy is a covariant measure of how much a fluid element has expanded or contracted. For this reason we will make extensive use of entropy density of fluid and express the EoS in terms of it, *i.e.* $\rho[s, \ldots]$ and $p[s, \ldots]$ rather than $p[\rho, \ldots]$.

# III. Curvature Generated Isocurvature Generalities

In this section we estimate the amplitude of isocurvature inhomogeneities that may be generated by curvature inhomogeneities. The analysis only derives consequences of the requirement that local isocurvature inhomogeneities depend only on the local curvature inhomogeneities. The specifics of the physics producing the isocurvature is not discussed. Only generic predictions are given which are not very quantitative. Unlike in the rest of this paper a stochastic model of both curvature and generation isocurvature is described in terms probability distribution functions (pdfs). This general treatment includes deterministic models as a special case where the pdfs are $\delta$-function

## IIIa Stochastic Generation of Isocurvature from Stochastic Curvature

Here we define an isocurvature inhomogeneity as occurring whenever a covariant scalar quantity, $\mathcal{I}$, defined by the matter (not the velocities) is not constant on surfaces of constant $\rho$. The simplest case is $\mathcal{I} = p$. If $p$ is not constant on surfaces of constant $\rho$ the fluid is abarotropic which is a type of isocurvature inhomogeneity. Here we assume an initial state (after inflation) which is barotropic, irrotational ($\omega_{\alpha\beta} = 0$) and a perfect fluid ($q_{\alpha\beta} = 0$) and with no nonzero isocurvature inhomogeneities whatsoever. Curvature inhomogeneities are assumed, *i.e.* where $\theta$ is not constant on surfaces of constant $\rho$. This initial state is what is meant by pure curvature initial conditions. With this initial state, assumed to persist for some time, a matter (or space-time) 4-velocity, $u^\alpha$, a density $\rho$, and an equation-of-state, $p[\rho]$. These ingredients define an entropy density $s$ (§IId) and the kurvature $K$ (§IIb2). Since $q_{\alpha\beta} = 0$, on super-horizon scales where pressure gradients may be neglected $\dot{K} = -\frac{2}{3}\theta K$ so $K \propto s^{-2/3}$. Initially where all relevant scales are outside the horizon one can define a quantity $\mathcal{K} \propto s^{-2/3} K$ for each fluid element which is initially constant in time. $\mathcal{K}$ like $K$ and $s$ is a local covariant quantity. Assume the initial spatial distribution of $\mathcal{K}$ is generated by a stochastic process. Consider two causally disconnected fluid elements 1 and 2 with $\mathcal{K}$ values $\mathcal{K}_1$ and $\mathcal{K}_2$. The stochastic process determines the joint pdf: $p^\times_\mathcal{K}[\mathcal{K}_1, \mathcal{K}_2]$ (× for cross correlation). If the process is homogeneous and isotropic then $p^\times_\mathcal{K}$ is symmetric in it's arguments and otherwise depends only on the initial separation of the two fluid elements. Also define the 1-point marginal pdf

$$p_K[K] \equiv \int_{-\infty}^{\infty} dK' \, p_K^\times[K, K']. \tag{9}$$

Only if the correlations are acausal will $p_K^\times[K_1, K_2] \neq p_K[K_1] p_K[K_2]$ which we assume is the case.

Consider a local covariant quantity, $\mathcal{I}$, which initially depends only on the local $\rho$ but due to curvature inhomogeneities in K may evolve inhomogeneously generating isocurvature inhomogeneities. The local value $\mathcal{I}$ can depend not only K along the same streamline but also on K in a nearby region of fluid elements but a region no larger than the causal horizon since inflation. Stochastic generation of $\mathcal{I}$ in may be expressed by the conditional pdf $p_\mathcal{I}[\mathcal{I} \mid K_{\text{reg}}]$ where $K_{\text{reg}}$ represents K field in this causal region. Here it is supposed that the $K_{\text{reg}}$ dependence is only a dependence on the average $\overline{K}$ in a region. Since spatial averages are more strongly correlated at large separations than average spatial gradients or higher order spatial moments it is likely that the dependence on an average $\overline{K}$ will dominate the covariance of $\mathcal{I}_1$ and $\mathcal{I}_2$. The spatial extent of the averaging region depends on local physics and is often not as large as the causal horizon. In most cases it is bounded by the smaller sound horizon. In the freeze-in analysis below we find this is the case. The 1-point marginal pdf of $\mathcal{I}$ is

$$p_\mathcal{I}[\mathcal{I}] = \int_{-\infty}^{\infty} d\overline{K} \, p_{\overline{K}}[\overline{K}] \, p_\mathcal{I}[\mathcal{I} \mid \overline{K}]. \tag{10}$$

.

so the joint pdf of $\mathcal{I}$ at 2-points on the constant $\rho$ hypersurface is

$$p_\mathcal{I}^\times[\mathcal{I}_1, \mathcal{I}_2] = \int_{-\infty}^{\infty} d\overline{K}_1 \, p_\mathcal{I}[\mathcal{I}_1 \mid \overline{K}_1] \int_{-\infty}^{\infty} d\overline{K}_2 \, p_\mathcal{I}[\mathcal{I}_2 \mid \overline{K}_1] \, p_{\overline{K}}^\times[\overline{K}_1, \overline{K}_2] \tag{11}$$

which like $p_K^\times$ is symmetric in it's argument and otherwise depends only on the initial separation of the two streamlines. Here $p_{\overline{K}}^\times$ is the convolution of $p_K^\times$ with the spatial window function describing the spatial average.

If there are no correlations between K in the two spatial regions then $p_K^\times[K, K'] = p_K[K] p_K[K']$ so $p_{\overline{K}}^\times[\overline{K}_1, \overline{K}_2] = p_{\overline{K}}[\overline{K}_1] p_{\overline{K}}[\overline{K}_2]$, and $p_\mathcal{I}^\times[\mathcal{I}_1, \mathcal{I}_2] = p_\mathcal{I}[\mathcal{I}_1] p_\mathcal{I}[\mathcal{I}_2]$. Even in this case if $\mathcal{I}$ evolves stochastically then $\mathcal{I} \neq \mathcal{I}[\rho]$ randomly and there will still be uncorrelated isocurvature inhomogeneities *not* generated from the curvature inhomogeneities. For deterministic evolution there will be no isocurvature inhomogeneities where there are no curvature inhomogeneities, *i.e.* $\mathcal{I} = \mathcal{I}[\rho]$ unless $p_K^\times[K, K'] \neq p_K[K] p_K[K']$.

## IIIb Covariance Functions

One can quantify the amplitude of the causal correlations by the normalized covariance

$$C_{\overline{K}}^\times \equiv \frac{\langle \overline{K}_1 \overline{K}_2 \rangle - \langle \overline{K} \rangle^2}{\langle \overline{K}^2 \rangle} \qquad C_\mathcal{I}^\times \equiv \frac{\langle \mathcal{I}_1 \mathcal{I}_2 \rangle - \langle \mathcal{I} \rangle^2}{\langle \mathcal{I}^2 \rangle} \tag{12}$$

where $\langle \cdots \rangle$ indicating an average over realizations so

$$\begin{aligned}
\langle \overline{K}_1^2 \rangle = \langle \overline{K}_2^2 \rangle = \langle \overline{K}^2 \rangle &= \int_{-\infty}^{\infty} d\overline{K} \, \overline{K}^2 \, p_{\overline{K}}[\overline{K}] \\
\langle \overline{K}_1 \overline{K}_2 \rangle &= \int_{-\infty}^{\infty} d\overline{K}_1 \, \overline{K}_1 \int_{-\infty}^{\infty} d\overline{K}_2 \, \overline{K}_2 \, p_{\overline{K}}^\times[\overline{K}_1, \overline{K}_2] \\
\langle \mathcal{I}_1^2 \rangle = \langle \mathcal{I}_2^2 \rangle = \langle \mathcal{I}^2 \rangle &= \int_{-\infty}^{\infty} d\mathcal{I} \, \mathcal{I}^2 \int_{-\infty}^{\infty} d\overline{K} \, p_\mathcal{I}[\mathcal{I} \mid \overline{K}] \, p_{\overline{K}}[\overline{K}] \\
\langle \mathcal{I}_1 \mathcal{I}_2 \rangle &= \int_{-\infty}^{\infty} d\mathcal{I}_1 \, \mathcal{I}_1 \int_{-\infty}^{\infty} d\overline{K}_1 \, p_\mathcal{I}[\mathcal{I}_1 \mid \overline{K}_1] \int_{-\infty}^{\infty} d\mathcal{I}_2 \, \mathcal{I}_2 \int_{-\infty}^{\infty} d\overline{K}_2 \, p_\mathcal{I}[\mathcal{I}_2, \overline{K}_2] \, p_{\overline{K}}^\times[\overline{K}_1, \overline{K}_2]
\end{aligned} \tag{13}$$

and therefore

$$C_\mathcal{I}^\times = \frac{\int_{-\infty}^{\infty} d\mathcal{I}_1 \, \mathcal{I}_1 \int_{-\infty}^{\infty} d\mathcal{I}_2 \, \mathcal{I}_2 \int_{-\infty}^{\infty} d\overline{K}_1 \, p_\mathcal{I}[\mathcal{I}_1 \mid \overline{K}_1] \int_{-\infty}^{\infty} d\overline{K}_2 \, p_\mathcal{I}[\mathcal{I}_2, \overline{K}_2] \left( p_{\overline{K}}^\times[\overline{K}_1, \overline{K}_2] - p_{\overline{K}}[\overline{K}_1] p_{\overline{K}}[\overline{K}_2] \right)}{\int_{-\infty}^{\infty} d\mathcal{I} \, \mathcal{I}^2 \int_{-\infty}^{\infty} d\overline{K} \, p_\mathcal{I}[\mathcal{I} \mid \overline{K}] \, p_{\overline{K}}[\overline{K}]} \tag{14}$$

If $\overline{K}_1$ and $\overline{K}_2$ are uncorrelated, $p_{\overline{K}}^\times[\overline{K}_1, \overline{K}_2] = p_{\overline{K}}[\overline{K}_1] p_{\overline{K}}[\overline{K}_2]$ and $C_\mathcal{I}^\times = C_{\overline{K}}^\times = 0$. The only dependence of $C_\mathcal{I}$ on the pair of fluid elements is through $p_{\overline{K}}^\times[\overline{K}_1, \overline{K}_2]$ which for statistically homogeneous isotropic initial states only depends on the initial separation which we denote below by $r$ without a rigorous definition.

K is 1st order in deviations from homogeneity and isotropy so in the simplest models one would expect the distribution of both K and of $\overline{K}$ to be Gaussian random noise. In inflationary cosmology the post inflation background cosmology has very small average curvature so Gaussian random noise will have zero mean : $(\langle \overline{K} \rangle = \langle K \rangle = 0)$. The joint and marginal pdf are

$$p_{\bar{K}}^{\times}[\bar{K}_1, \bar{K}_2] = \frac{e^{-\frac{\bar{K}_1^2 + \bar{K}_2^2 - 2\bar{K}_1\bar{K}_2 C_{\bar{K}}}{2(1-C_{\bar{K}}^2)\langle\bar{K}^2\rangle}}}{2\pi\sqrt{1-C_{\bar{K}}^2}\,\langle\bar{K}^2\rangle} \qquad p_{\bar{K}}[\bar{K}] = \frac{e^{-\frac{\bar{K}^2}{2\langle\bar{K}^2\rangle}}}{\sqrt{2\pi\langle\bar{K}^2\rangle}} \quad . \tag{15}$$

and the normalized covariance is

$$C_{\mathcal{I}}^{\times} = \frac{\int_{-\infty}^{\infty} d\mathcal{I}_1\,\mathcal{I}_1 \int_{-\infty}^{\infty} d\mathcal{I}_2\,\mathcal{I}_2 \int_{-\infty}^{\infty} d\bar{K}_1\, p_{\mathcal{I}}[\mathcal{I}_1 | \bar{K}_1]\, p_{\bar{K}}[\bar{K}_1] \int_{-\infty}^{\infty} d\bar{K}_2\, p_{\mathcal{I}}[\mathcal{I}_2 | \bar{K}_2]\, p_{\bar{K}}[\bar{K}_2] \left( \frac{e^{-\frac{C_{\bar{K}}^2(\bar{K}_1^2 + \bar{K}_2^2) - 2\bar{K}_1\bar{K}_2 C_{\bar{K}}}{2(1-C_{\bar{K}}^2)\langle\bar{K}^2\rangle}}}{\sqrt{1-C_{\bar{K}}^2}} - 1 \right)}{\int_{-\infty}^{\infty} d\mathcal{I}\,\mathcal{I}^2 \int_{-\infty}^{\infty} d\bar{K}\, p_{\mathcal{I}}[\mathcal{I} | \bar{K}]\, p_{\bar{K}}[\bar{K}]} \tag{16}$$

For Gaussian statistics the only dependence of $C_{\mathcal{I}}$ on the pair of fluid elements is through $C_{\bar{K}}$. If the $\mathcal{I}$ distribution does not depend on $\bar{K}$, i.e. if $p_{\mathcal{I}}[\mathcal{I} | \bar{K}] = p_{\mathcal{I}}[\mathcal{I}]$, then $C_{\mathcal{I}} = 0$.

## IIIb Large Scale Isocurvature Bias

In most cases of interest kurvature correlations, $|C_{\bar{K}}|$, decreases rapidly with separation, *i.e.* if the separation is large then $C_{\bar{K}}$ is small. For $|C_{\bar{K}}| \ll 1$ one can approximate $C_{\mathcal{I}}$ by the first few terms in a Taylor expansion about $C_{\bar{K}} = 0$ or more particularly the term in parentheses. All terms at any order in this are a multinomial in $\bar{K}_1$, $\bar{K}_2$ and $C_{\bar{K}}$. The integral of each term in the multinomial thus factorizes into separate integrals for each of the two fluid elements, These separate integrals are 1-point statistics. These 1-point statistics are determined by *local* physics of the matter model not on super-horizon inhomogeneities. Explicitly the expansion is

$$C_{\mathcal{I}}^{\times} = B_1\, C_{\bar{K}}^{\times} + B_2\, C_{\bar{K}}^{\times 2} + B_3\, C_{\bar{K}}^{\times 3} + \cdots \tag{17}$$

This expansion is similar to though not exactly analogous to a bias expansion of galaxy clustering. Super-horizon curvature inhomogeneities are encoded in $C_{\bar{K}}^{\times}$ while the local physics is encoded in the "bias" factors $B_i$ which are products of these 1-point statistics. The bias factor are just constants in the sense that they are independent of the separation of the two fluid elements. The 1st two bias factors are

$$B_1 = r_{\bar{K}\mathcal{I}}^2 \qquad B_2 = \frac{1}{2} \frac{\left(\langle\bar{K}^2\,\mathcal{I}\rangle - \langle\bar{K}^2\rangle\langle\mathcal{I}\rangle\right)^2}{\langle\bar{K}^2\rangle^2 \langle\mathcal{I}^2\rangle} \quad \text{where} \quad r_{\bar{K}\mathcal{I}} = \frac{\langle\bar{K}\,\mathcal{I}\rangle}{\sqrt{\langle\bar{K}^2\rangle\langle\mathcal{I}^2\rangle}} \quad . \tag{18}$$

Note that $B_1$, $B_2 \geq 0$ and $-1 \leq r_{\bar{K}\mathcal{I}} \leq 1$ so $0 \leq B_1 \leq 1$.

When $|C_{\bar{K}}^{\times}| \ll 1$ it is the first (linear) bias term which is likely to determine $C_{\mathcal{I}}^{\times}$. That is so long as $\langle\bar{K}\,\mathcal{I}\rangle \neq 0$. If one neglects non-linear bias then $0 \leq C_{\mathcal{I}}^{\times}/C_{\bar{K}}^{\times} \leq 1$. Thus in terms of the normalized covariances super-horizon correlations in $\mathcal{I}$ are never larger than those of $\bar{K}$.

## IIIb Yield Correlation Functions

Now consider the case where, by definition, $\mathcal{I} \geq 0$. For this case use the notation $\mathcal{I} \to Y$. $Y$ might be the *yield* or number density of some species of matter. Since $Y$ is defined on a surface of constant $\rho$ it is a local covariant quantity. Another local covariant quantity which used extensively in §IV is the fractional overdensity $\Upsilon \equiv (Y - \langle Y \rangle)/\langle Y \rangle$. A correlation statistic conventionally used in cosmology to describe positive quantities such as yield is

$$\xi \equiv \frac{\langle Y_1\, Y_2\rangle - \langle Y \rangle^2}{\langle Y \rangle^2} = \langle \Upsilon_1\, \Upsilon_2 \rangle = \frac{\langle Y^2 \rangle}{\langle Y \rangle^2}\, C_Y^{\times} \tag{19}$$

For microscopic production of large numbers particles the yield will not vary much between different regions so $\langle Y^2 \rangle \cong \langle Y \rangle^2$ and $\xi \cong C_Y$. In the case of linear bias $\xi \cong r_Y^2\, C_{\bar{K}}$ so $\xi/C_{\bar{K}} \leq 1$ when $\langle Y^2 \rangle \cong \langle Y \rangle^2$. At the other extreme one can apply this to the production of macroscopic an rare objects. Primordial black holes comes to mind since to avoid overproduction of black holes the probability of producing even one black hole in a causal patch most be extremely small. In such cases $\langle Y^2 \rangle \gg \langle Y \rangle^2$ and if $\langle \bar{K}\,\Upsilon \rangle$ is not to small one can have $\xi/C_{\bar{K}} \gg 1$. Unlike for the $C_Y^{\times}$ measure, for the $\xi$ measure of compositional isocurvature amplitude of for rare objects can be arbitrarily larger than $C_{\bar{K}}$!

## IIIb The Isocurvature Bump

Other explicit expression for $\xi$ are

$$\xi = \frac{\langle \overline{K}\, Y \rangle^2}{\langle \overline{K}^2 \rangle \langle Y \rangle^2}\, C^{\times}_{\overline{K}} = \frac{\langle \overline{K}\, Y \rangle^2}{\langle \overline{K}^2 \rangle}\, C^{\times}_{\overline{K}} = \frac{\langle \overline{K}\, Y \rangle^2 \langle \overline{K}_1\, \overline{K}_2 \rangle}{\langle \overline{K}^2 \rangle^2} \tag{20}$$

where we have used $\langle \overline{K} \rangle = 0$.

On accessible scales the initial isocurvature inhomogeneities appear to have close to a scale invariant (Harrison-Zel'dovich) spectrum. For this spectrum the gravitational potential correlation only fall off logarithmically. That being he case eq. 6 tells us that $\langle K_1\, K_2 \rangle \propto r^{-4}$. The coefficient of this proportionality depends on the details of spatial averaging which depends on the matter model. On scales much larger than the averaging scale

$$\xi \approx \frac{\langle \overline{K}\, Y \rangle^2}{\langle \overline{K}^2 \rangle} \left( \frac{r_w}{r} \right)^4 \qquad r \gg r_w \tag{21}$$

where $r_w$ is a characteristic averaging radius. Averaging will suppress correlations at smaller separations. This suppression can be sufficient to insure $\xi < \infty$ as $r \to 0$ as found in the freeze-in analysis below. Given the rapid fall-off for large $r$ one can reasonably say that the isocurvature correlations are primarily generated on the averaging scale $\sim r_w$. In terms of power spectra one could then describe the spectrum of isocurvature modes generated as having a "bump" around wavenumbers $|\boldsymbol{k}| \sim 1/r_w$. This is indeed what is found for dark matter freeze-in treated next.

# IV. Dark Matter Freeze-in

In freeze-in models of dark matter it is assumed to be very feebly coupled to SM matter and at some very early time is assumed to have negligible density. The feeble coupling however allows some production of DM as the universe expands and usually most the DM is produced during a short cosmological *epoch of freeze-in*. DM production comes to an end at some redshift usually when the SM temperature falls below the mass of some DM particle necessary for this reaction to proceed. Often this particle is the dark matter particle itself. Eventually one or more massive DM particle(s) come to dominate the mass density of the universe as what we refer to as dark matter. Here it is assumed there is only one such species with mass $m_d$.

To quantify the effects of freeze-in define the *dark matter number yield* and *baryon asymmetry parameter*:

$$_n Y \equiv \frac{_d n}{_s s} \qquad _b \eta \equiv \ln\left[ \frac{_b n}{_s s} \right] \tag{22}$$

where $_d n$ is the dark matter particle density (particles *plus* anti-particles), $_s s$ the SM entropy density and $_b n$ the baryon number density (baryons *minus* anti-baryons). $_n Y$ starts out uniformly equal to zero but will grow during freeze-in. We assume that $_b \eta$ starts out with some uniform value $\overline{\eta}_{bi}$ and will also grows during freeze-in as SM entropy is lost by DM production. Baryon number is not changed but rather its abundance is enhanced relative to SM entropy which is lost. Inhomogeneous growth will lead to compositional inhomogeneities in $_n Y$ and aba inhomogeneities in $_b \eta$. When the temperature falls below $\sim 1$ GeV the remnant baryons from the baryon asymmetry becomes the only non-relativistic component of the SM matter and is sometimes treated separately. If one treats the remnant baryons as a separate fluid what would otherwise be treated as an abarotropic SM inhomogeneity could be treated as a compositional inhomogeneity of baryons relative to SM.

## IVa Separate Universes Modelling (SUM)

One simple way to analyze cosmic evolution on super-horizon scales is to treat each super-horizon patch as separate FLRW cosmology. This is usually justified by stating that different horizon patches are not causally connected and therefore do not communicate with each other. The differing values of quantities in different patches are referred to as "local" values and the different patches associated with separate universes are "localities". Localities differ in that they have differing values kurvature constant, $k$, which gives the kurvature $K[t]$ and different expansion laws $a[t]$. Dark matter freeze-in is modeled with the same chemical rate equations and the dark matter yield will vary between localities only due to the differing expansion laws. Quantities such as the DM and SM density, $_d\rho$ and $_s\rho$, are local covariant values. In this manner of separate universe modelling (SUM) one associates the distribution of local values of $_d\rho$ and $_s\rho$ with the distribution of inhomogeneities in these quantities in the actual inhomogeneous universe. These distributions will depend on the assumed distribution of $k$ which one associates with the distribution of curvature inhomogeneities in actual universe by eq. 6.

### IVa1 SUM Freeze-in

A salient feature of freeze-in is that the feeble coupling combined with low initial DM density means there is negligible absorption of DM by SM. Use $\dot{e}$ to represent the volumetric rate of energy transfer from SM to DM and $\dot{n}$ to represent the volumetric rate of production of dark matter particles. Since the SM remains in thermal equilibrium $\dot{e}$ and $\dot{n}$ depend only on the parameters describing the state of SM: $_s s$ and $_b \eta$. Since the baryon asymmetry is so small during freeze-in $_b \eta$ is treated as a "spectator field" meaning baryon number is conserved and it is henceforth assumed that DM production is insensitive to the value of $_b \eta$. Since the SM is assumed barotropic it's density and pressure are given by two functions of the local SM entropy and baryon density $_s s$: $\rho_s[_s s, _b \eta]$ and $p_s[_s s, _b \eta]$. Since $_s \dot{\rho} = -3 \frac{\dot{a}}{a} (_s \rho + _s p) - \dot{e}$ and $_s \dot{s} = _s \dot{\rho} \, _s s /(_s \rho + _s p)$ one finds that in a FLRW context DM production is described by

$$_s \dot{s}[t] = -3 \frac{\dot{a}[t]}{a[t]} \, _s s[t] - \frac{\dot{e}[_s s[t]]}{\rho_s[_s s[t], _b \eta[t]] + p_s[_s s[t], _b \eta[t]]} \, _s s[t]$$
$$_d \dot{\rho}[t] = -3 \frac{\dot{a}[t]}{a[t]} (_d \rho[t] + _d p[t]) + \dot{e}[_s s[t], _b \eta[t]]$$
$$_d \dot{n}[t] = -3 \frac{\dot{a}[t]}{a[t]} \, _d n[t] + \dot{n}[_s s[t], _b \eta[t]] \qquad (23)$$
$$_b \dot{n}[t] = -3 \frac{\dot{a}[t]}{a[t]} \, _b n[t]$$

The dark matter yield and baryon asymmetry evolve as

$$_n \dot{Y}[t] = \frac{\dot{n}[_s s[t]]}{_s s[t]} + \frac{\dot{e}[_s s[t]]}{\rho_s[_s s[t], _b \eta[t]] + p_s[_s s[t], _b \eta[t]]} \, _n Y[t] \qquad _b \dot{\eta}[t] = \frac{_b \dot{n}}{_b n} - \frac{_b \dot{s}}{_s s} = \frac{\dot{e}[_s s[t]]}{\rho_s[_s s[t], _b \eta[t]] + p_s[_s s[t], _b \eta[t]]} \quad . \qquad (24)$$

It is assumed that $\dot{n}$ and $\dot{e}$ falls off rapidly enough at large and small $s$ that these two quantities converge to finite values at late times: $_n Y[t] \to {_n Y_f}$ and $_b \eta[t] \to {_b \eta_f}$. Formal solutions for these asymptotic values are given by

$$_n Y_f = \int_0^\infty dt \frac{\dot{n}[_s s[t]]}{_s s[t]} \operatorname{Exp}\!\left[\int_t^\infty dt' \frac{\dot{e}[_s s[t']]}{\rho_s[_s s[t'], _b \eta[t']] + p_s[_s s[t'], _b \eta[t']]}\right] \qquad _b \eta_f = \bar{\eta}_{\text{bi}} + {_e Y_f} \qquad _e Y_f = \int_0^\infty dt \frac{\dot{e}[_s s[t]]}{\rho_s[_s s[t'], _b \eta[t']] + p_s[_s s[t'], _b \eta[t']]} \qquad (25)$$

These are only implicit solutions since they involve the unknown $_s s[t]$ and $_b \eta[t]$. The *baryon enhancement* is $_b \eta_f - \bar{\eta}_{\text{bi}}$ given by the fractional SM entropy loss $_e Y_f$.

One obtains different local values of $_n Y_f$ and $_e Y_f$ because different localities have different different expansion laws, $a[t]$, when they have different values of kurvature. One can express this using

$$\rho[t] \equiv {_s \rho}[_s s[t], _b \eta[t]] + {_d \rho[t]} \qquad K[t] \equiv \frac{8\pi G \tilde{\rho}[t]}{3} - \left(\frac{\dot{a}[t]}{a[t]}\right)^2 = \frac{k}{a[t]^2} \quad .$$

Changing the local clock variable from $t$ to $a$ using $dt = \frac{da}{\tilde{a}} \frac{1}{H[a]}$ where $H[a] = \frac{\dot{a}[t[a]]}{a}$ one obtains

$$_n Y_f = \int_0^\infty \frac{da}{a} \sqrt{_s \Omega[a]} \, \epsilon_n[_s s[a], _s \eta[a]] \operatorname{Exp}\!\left[\int_a^\infty \frac{da'}{a'} \sqrt{_s \Omega[a']} \, \epsilon_e[_s s[a'], _s \eta[a']]\right]$$
$$_e Y_f = \int_0^\infty \frac{da}{a} \sqrt{_s \Omega[a]} \, \epsilon_e[_s s[a]] \qquad (26)$$

where

$$_s \Omega[a] \equiv \frac{8\pi G \rho_s[_s s[a]]}{3 H[a]^2} = \frac{1}{1 + \frac{_d \rho[a]}{\rho_s[_s s[a]]} - \frac{3}{8\pi G \rho_s[_s s[a]]} \frac{k}{a^2}}$$

$$\epsilon_n[s] \equiv \sqrt{\frac{3}{8\pi G \rho_s[s]}} \, \frac{\dot{n}[s]}{s} \qquad \epsilon_e[s] \equiv \sqrt{\frac{3}{8\pi G \rho_s[s]}} \, \frac{\dot{e}[s]}{\rho_s[s] + p_s[s]} \quad . \qquad (27)$$

are dimensionless parameterizations of $k$, $\dot{n}$ and $\dot{e}$ respectively. Differing values of $k$ will lead to different values of $_n Y_f$ and $_e Y_f$.

### IVa2 Feeble Freeze-In Approximation

Freeze-in makes use of the feeble interaction of DM with SM to prevent the DM from thermalizing. While one can imagine intermediate scenarios of near thermalization here it is assumed that the DM remains far from thermalization. This generally requires that $\epsilon_e \ll 1$ densities and pressures are a small fraction of their SM counterparts during freeze-in. In this sense, during freeze-in the DM is also treated as a spectator field. The "feeble freeze-in approximation" is to expand quantities to lowest order in $\epsilon_e$.

$$_s s[a] = {_s s[a_{\text{fid}}]} \left(\frac{a_{\text{fid}}}{\tilde{a}}\right)^3 (1 + O[\epsilon_e]) \qquad _d \rho[a] = 0 + O[\epsilon_e] \qquad \rho[a] = {_s \rho[a]} + O[\epsilon_e] \qquad (28)$$

where $\tilde{a}_{\text{fid}}$ is any fiducial scale factor. To lowest order the SM is being treated as adiabatic. For adiabatic SM there is a simple relationship between $_s s$ and $a$ which facilitates the change of clock variables from the non-covariant $\tilde{a}$ to the covariant $_s s$:

$$_xY_f = \frac{1}{3} \int_0^\infty \frac{ds}{s} \frac{\epsilon_x[s]}{\sqrt{1-_s\kappa[s]}} + O[\epsilon_n \epsilon_e, \epsilon_e^2] \qquad _s\kappa[_ss] \equiv \frac{3K[s_{fid}]}{8\pi G \rho_s[s]} \left(\frac{_ss}{s_{fid}}\right)^{2/3} = \frac{_s\Omega[_ss]-1}{_s\Omega[_ss]} \qquad (29)$$

for $x$ = n, e, where $s_{fid}$ is any fiducial entropy density chosen to have the same value in all separate universes. Here $K \propto a^{-2} \propto {}_ss^{2/3}$ has been used. As indicated here the quantity $_s\kappa$ is just a measure of the *local* density parameter as $_s\Omega$.

### IVa3 Einstein deSitter SUM Freeze-In

If one applies SUM to a flat ($k = 0$) cosmology then $_s\kappa$ will fluctuate around zero. It is these fluctuations which give the inhomogeneities in $_xY_f$. Linear perturbation theory neglects terms quadratic and higher in $_s\kappa^2$ i.e.

$$_xY_f = {}_x\overline{Y}_f(1 + {}_x\Upsilon_f) + O[\epsilon_n\epsilon_e, \epsilon_e^2, \epsilon_n {}_s\kappa^2, \epsilon_e {}_s\kappa^2] \qquad _x\overline{Y}_f = \frac{1}{3}\int_0^\infty \frac{ds}{s} \epsilon_x[s] \qquad _x\Upsilon_f = \frac{1}{2}\frac{\int_0^\infty \frac{ds}{s} \epsilon_x[s] {}_s\kappa[s]}{\int_0^\infty \frac{ds}{s} \epsilon_x[s]} \qquad (30)$$

for $x$ = n, e. $_n\overline{Y}_f$ and $_e\overline{Y}_f$ do not depend on $_s\tilde{\kappa}$, i.e. are the same for all separate universes, and are the mean values of the dark matter yield and energy transfer and hence the tilde has been exchanged for an bar. $_x\Upsilon_f$ gives the fractional inhomogeneity in the dark matter yield and is proportional to $_s\kappa[_ss]$ which is proportional to the the kurvature. $_e\overline{Y}_f$ gives the shift in the baryon-to-photon ratio, $_b\eta$, and $_e\Upsilon_f$ gives the fractional inhomogeneity in $_b\eta$. Eqs 30 encapsulate the main new results of this paper. Observational consequences will be explored below.

## IVb Super-horizon Feeble Freeze-In Revisited

Here the SUM result is re-derived in an inhomogeneous FLRW setting but using a super-horizon approximation. One can re-express the feeble freeze-in derivation in terms of perturbations from a single FLRW space-time. Only the dark matter yield is considered here as the baryon enhancement follows similarly. Settling on the notion that super-horizon means neglecting pressure gradients the DM will remain comoving with SM on super-horizon scales when the differing pressure gradients are neglected. The two fluids thus have a common rate of expansion, $_s\theta = {}_d\theta = \theta$ and the convective proper time derivative of the dark matter yield is

$$_n\dot{Y} \cong \frac{d\dot{n}}{_ss} - \frac{dn}{_ss} \frac{_s\dot{s}}{_ss} = {}_nY({}_s\theta - {}_d\theta) + \Gamma_n[{}_ss, {}_b\eta] - {}_nY\Gamma_e[{}_ss, {}_b\eta] = \Gamma_n[{}_ss, {}_b\eta] - {}_nY\Gamma_e[{}_ss, {}_b\eta] \qquad _b\dot{\eta} = \Gamma_e[{}_ss, {}_b\eta]\ {}_n$$

$$\Gamma_n[{}_ss, {}_b\eta] \equiv \frac{\dot{n}[{}_ss, {}_b\eta]}{_ss} \qquad \Gamma_e[{}_ss, {}_b\eta] \equiv \frac{\dot{n}[{}_ss, {}_b\eta]}{\rho_s[{}_ss, {}_b\eta] + p_s[{}_ss, {}_b\eta]} \qquad (31)$$

which is just eq 24. The feeble freeze-in approximation amounts to neglecting the 2nd term in $_n\dot{Y}$ and ignoring the $_b\eta$ dependence since $_b\eta$ is approximated as unchanged for feeble freeze-in. One finds for the final yields, $x$ = n, e,

$$_xY_f \cong \int_0^\infty dt\ \Gamma_x[{}_ss] \qquad (32)$$

In the feeble freeze-in approximation SM remains dominant and the non-adiabaticity of SM from DM production is assumed negligible so $\dot{s} \cong -s\theta$. Re-express this for small inhomogeneous perturbations from a uniform flow: $_ss[t, \boldsymbol{x}] = {}_s\overline{s}[t] + {}_s\delta s[t, \boldsymbol{x}]$ and $\theta[t, \boldsymbol{x}] = \overline{\theta}[t] + \delta\theta[t, \boldsymbol{x}]$ where $_s\dot{\overline{s}} = -\overline{\theta}\ {}_s\overline{s}$ and $_s\dot{\delta s} = -(\overline{\theta}\ {}_s\delta s + \delta\theta\ {}_s\overline{s}) + O[2]$. Essentially these are equations in a synchronous comoving gauge. Defining $\Gamma_x'[s] \equiv \frac{d\Gamma_x[s]}{ds}$ one finds $\frac{d}{dt}\Gamma_x[\overline{s}] = -\overline{s}\ \overline{\theta}\ \Gamma_x'[\overline{s}]$ or $\Gamma_x'[\overline{s}] = -\frac{1}{\overline{s}\ \overline{\theta}}\frac{d}{dt}\Gamma_x[\overline{s}]$. Taylor expanding for small $\delta s$ and integrating by parts assuming $\Gamma_x$ goes to zero at both early and late times

$$_xY_f[\boldsymbol{x}] \cong \int_0^\infty dt\ \Gamma_x[{}_ss[t, \boldsymbol{x}]] \cong \int_0^\infty dt\ \Gamma_x[{}_s\overline{s}[t]] + \int_0^\infty dt\ {}_s\delta s[t, \boldsymbol{x}]\ \Gamma_x'[{}_s\overline{s}[t, \boldsymbol{x}]]\ .$$

$$= \int_0^\infty dt\ \Gamma_x[{}_s\overline{s}[t]] - \int_0^\infty dt\ \frac{_s\delta s[t, \boldsymbol{x}]}{_s\overline{s}[t]\ \overline{\theta}[t]}\frac{d}{dt}\Gamma_x[{}_s\overline{s}[t]]$$

$$= \int_0^\infty dt\ \Gamma_x[{}_s\overline{s}[t]] + \int_0^\infty dt\ \frac{\Gamma_x[{}_s\overline{s}[t]]}{\overline{\theta}[t]}\left(\frac{_s\dot{\delta s}[t, \boldsymbol{x}]}{_s\overline{s}[t]} - \frac{_s\dot{\overline{s}}[t]}{_s\overline{s}[t]}\frac{_s\delta s[t, \boldsymbol{x}]}{_s\overline{s}[t]} - \frac{\dot{\overline{\theta}}[t]}{\overline{\theta}[t]}\frac{_s\delta s[t, \boldsymbol{x}]}{_s\overline{s}[t]}\right) \qquad (33)$$

$$= \int_0^\infty dt\ \Gamma_x[{}_s\overline{s}[t]]\left(1 - \frac{\delta\theta[t, \boldsymbol{x}]}{\overline{\theta}[t]} - \frac{\dot{\overline{\theta}}[t]}{\overline{\theta}[t]^2}\frac{_s\delta s[t, \boldsymbol{x}]}{_s\overline{s}[t]}\right)$$

For adiabatic fluids $d\rho = (\rho + p) d\ln[s]$ requiring $\frac{\delta s}{s} = \frac{1}{1+\overline{w}}\frac{\delta\rho}{\rho}$ where $\overline{w} \equiv \frac{\overline{p}}{\overline{\rho}}$. For the flat FLRW background cosmology assumed here $\dot{\overline{\theta}} = -\frac{1}{2}(1 + \overline{w})\overline{\theta}^2$ so

$$_xY_f[\boldsymbol{x}] \cong \int_0^\infty dt \, _x\Gamma[\overline{s}[t]] \left(1 - \frac{\delta\theta[t, \boldsymbol{x}]}{\overline{\theta}[t]} + \frac{1}{2} \frac{_s\delta\rho[t, \boldsymbol{x}]}{_s\overline{\rho}[t]}\right). \tag{34}$$

For a flat FLRW space-time $\overline{\rho} = \frac{\overline{\theta}^2}{24\pi G}$ so $\overline{K} = \frac{8\pi G \overline{\rho}}{3} - \frac{1}{9}\overline{\theta}^2 = 0$ and one finds

$$K[t, \boldsymbol{x}] \cong \frac{8\pi G \delta\rho}{3} - \frac{2}{9}\delta\theta\,\overline{\theta} + O[2] = \frac{8\pi G \overline{\rho}}{3}\left(\frac{\delta\rho}{\overline{\rho}} - 2\frac{\delta\theta}{\overline{\theta}}\right) \tag{35}$$

so using the previously defined quantity $\kappa = \frac{3K}{8\pi G \rho} = \frac{\delta\rho}{\overline{\rho}} - 2\frac{\delta\theta}{\overline{\theta}}$ and once again assuming SM dominates during freeze in ($\overline{\rho} \cong {}_s\overline{\rho}$ and $\delta\rho \cong {}_s\delta\rho$) one finds

$$_xY_f[\boldsymbol{x}] \cong \int_0^\infty dt \, \Gamma_x[\overline{s}[t]]\left(1 + \frac{1}{2}\kappa[t, \boldsymbol{x}]\right). \tag{36}$$

Using $\Gamma_x[{}_s\overline{s}] = \frac{1}{3}\overline{\theta}\,\epsilon_x[{}_s\overline{s}] = -\frac{1}{3}\frac{{}_s\dot{\overline{s}}}{{}_s\overline{s}}\epsilon_x[{}_s\overline{s}]$ and taking the interval $t \in [0, \infty]$ to correspond to the interval ${}_s\overline{s} \in [\infty, 0]$ one finds

$$_xY_f[\boldsymbol{x}] \cong {}_x\overline{Y}_f\left(1 + {}_x\tilde{Y}_f[\boldsymbol{x}]\right) \qquad _x\overline{Y}_f = \frac{1}{3}\int_0^\infty \frac{ds}{s}\epsilon_x[s] \qquad _x\tilde{Y}_f[\boldsymbol{x}] \cong \frac{1}{2}\frac{\int_0^\infty \frac{ds}{s}\epsilon_x[s]\kappa[s,\boldsymbol{x}]}{\int_0^\infty \frac{ds}{s}\epsilon_x[s]} \tag{37}$$

where $\kappa[{}_s\overline{s}, \boldsymbol{x}] = \frac{3K[\overline{t}[{}_s\overline{s}],\boldsymbol{x}]}{8\pi G \rho_s[{}_s\overline{s}]}$ and $\overline{t}[{}_s\overline{s}]$ is the inverse function of ${}_s\overline{s}[t]$. This expression is the same as eq. 30 obtained by SUM. The local yield, which is the ratio of ${}_dn$ to ${}_ss$, is a local covariant quantity and the local final yield is time-independent. A covariant perturbative quantity like ${}_x\delta Y_f = {}_xY_f - {}_x\overline{Y}_f$ will only depend on the time-slicing (gauge) if ${}_x\overline{Y}_f$ is time dependent. Since ${}_x\overline{Y}_f$ is just a fixed number the final value, ${}_x\tilde{Y}_f$, is gauge invariant.

## IVc Analysis of Feeble Freeze-In

### IVc1 Super-horizon Feeble Freeze-In and Curvature Potential

The reader may be more familiar with gauge invariant perturbation theory using conformal time $\eta$ where $dt = a\,d\eta$. For perturbations from a flat FLRW space-time one may Fourier decompose the inhomogeneities, $f[\eta, \boldsymbol{x}] = \overline{f}[\eta] + \sum_{\boldsymbol{k}}\tilde{f}[\eta, \boldsymbol{k}]\,e^{i\boldsymbol{k}\cdot\boldsymbol{x}}$ where $\boldsymbol{k}$ is the comoving wavenumber. Using

$$\tilde{\kappa}[\eta, \boldsymbol{k}] = \frac{\tilde{\delta\rho}[\eta, \boldsymbol{k}]}{\overline{\rho}[\eta]} - 2\frac{\tilde{\delta\theta}[\eta, \boldsymbol{k}]}{\overline{\theta}[\eta]} = -\frac{2}{3}\frac{|\boldsymbol{k}|^2}{\mathcal{H}[\eta]^2}\tilde{\mathcal{R}}[\eta, \boldsymbol{k}] \tag{38}$$

where $\mathcal{H}[\eta] \equiv a'[\eta]/a[\eta]$ and $\tilde{\mathcal{R}}$ is the curvature potential As previously mentioned on super-horizon scales $\tilde{\mathcal{R}}[\eta, \boldsymbol{k}] = \tilde{\mathcal{R}}_0[\boldsymbol{k}]$ is a constant so the super-horizon final yields for $x = $ n, e are

$$_xY_f[\boldsymbol{x}] = \overline{Y}_f\left(1 + \sum_{\boldsymbol{k}} {}_x\tilde{Y}_f[\boldsymbol{k}]\,e^{i\boldsymbol{k}\cdot\boldsymbol{x}}\right) \quad _x\overline{Y}_f = \frac{1}{3}\int_0^\infty \frac{ds}{s}\epsilon_x[s] \quad _x\tilde{Y}_f[\boldsymbol{k}] \cong -\frac{1}{3}\tilde{\mathcal{R}}_0[\boldsymbol{k}]\left(\frac{|\boldsymbol{k}|}{\mathcal{H}_x^{fi}}\right)^2 \quad \mathcal{H}_x^{fi} \equiv \sqrt{\frac{\int_0^\infty \frac{ds}{s}\epsilon_x[s]}{\int_0^\infty \frac{ds}{s}\frac{\epsilon_x[s]}{\mathcal{H}[\eta[s]]^2}}} \quad |\boldsymbol{k}| \ll \mathcal{H}_x^{fi} \tag{39}$$

where $\eta[{}_s\overline{s}]$ is the inverse function of ${}_s\overline{s}[\eta]$. Here $\mathcal{H}_x^{fi}$ are characteristic value of $\mathcal{H}[\eta]$ during freeze-in which one can calculate for any freeze-in model which determines $\epsilon_n[{}_ss]$ and $\epsilon_e[{}_ss]$. The quantities ${}_x\tilde{Y}_f$ will be called *entropy ratio anomalies* since the default (non-anomalous) assumption is that the dark matter is produced primordially and the ratio of the number densities of baryons and the dark matter to the entropy SM entropy density are uniform. In the non-anomalous case ${}_x\tilde{Y}_f = 0$. These approximate expression for the anomalies are the main result of this paper.

### IVc2 Characteristic Temperatures

The $\mathcal{H}$'s have dimensions of inverse conformal time or equivalently comoving wavenumber. Pressure gradient induced motion will have a large effects on freeze-in yield on scales comparable or smaller than the sound horizon during freeze-in. Typically the SM ultra-relativistic during freeze-in and the comoving sound horizon is $\sqrt{3}/\mathcal{H}$ and the super-horizon approximation used is only valid for $|\boldsymbol{k}| \ll \mathcal{H}_x^{fi}$. Thus $\mathcal{H}_n^{fi} \approx \mathcal{H}_e^{fi}$ is roughly the limiting wavenumber for the results derived.

Here it is assumed that that the SM is nearly adiabatic, not only during feeble freeze-in, but also afterwards until the present epoch *i.e.* $_s\bar{s} \propto (1+z)^3$. For an SM model one can relate the entropy density to temperature, $s_s[T]$ and it may be more useful to re-express freeze-in physics in terms of temperature. To do this one can use the thermodynamic formulae

$$s_s[_\gamma \bar{T}] = \frac{2\pi^2}{45} g_s[_\gamma \bar{T}] _\gamma \bar{T}^3 \qquad \rho_s[_\gamma \bar{T}] = \frac{\pi^2}{30} g_\rho[_\gamma \bar{T}] _\gamma \bar{T}^4 \tag{40}$$

where $g_s$ and $g_\rho$ are effective number of degrees of freedom. The mean photon temperature, $_\gamma \bar{T}$, is specified since after $e^\pm$ annihilation the SM is described by separate photon and neutrino temperatures, $_\nu \bar{T} \neq {}_\gamma \bar{T}$ and one must distinguish the two. Even where a fluid is not in thermodynamic equilibrium $g_s$ and $g_\rho$ provide a useful parameterizations of entropy and energy density. With the $_\gamma T$ parameterization define $\epsilon_x[_\gamma T] = \epsilon_x[s_s[_\gamma T]]$. Specifically consider the case where freeze-in happens predominantly when the SM fluid maintains an ultra-relativistic EoS in which case $g_s[_\gamma T] \cong g_s^{fi}$ and $g_\rho[_\gamma T] \cong g_\rho^{fi}$ are approximately constant during freeze-in. In this case one can define characteristic temperatures for freeze-in which is linearly related to the characteristic wavenumbers:

$$T_x^{fi} \cong \sqrt{\frac{\int_0^\infty \frac{dT}{T} \epsilon_x[T]}{\int_0^\infty \frac{dT}{T^3} \epsilon_x[T]}} \qquad \mathcal{H}_x^{fi} \cong \sqrt{\frac{8\pi^3}{90} g_\rho^{fi}} \left(\frac{g_s^{eq}}{g_s^{fi}}\right)^{1/3} \frac{\sqrt{G}\, T_{eq}\, T_x^{fi}}{1+z_{eq}}. \tag{41}$$

Here $_\gamma \bar{T} = T_{eq}$ at matter-radiation equality at redshift $z = z_{eq}$. To obtain $\mathcal{H}_x^{fi}$ use $\frac{8\pi}{3} G\, {}_s\rho \cong (1+z)^2 \mathcal{H}^2$ and $(1+z_{eq})^3 \, _s s \cong (1+z)^3 \, _s s_s^{eq}$.

### IVc3 Concordance Cosmology Freeze-In

For an ultra-relativistic weakly interacting gas where all species have the same kinetic temperature a boson contributions 1 to both $g_s$ and $g_\rho$ for each helicity state while a fermion contributes 7/8. Baryons have negligible entropy but are included in $_s\rho$ as part of the definition of SM. For two temperature intervals of interest are

$$\begin{aligned} g_s[_\gamma \bar{T}] &\cong \frac{43}{11} & g_\rho[_\gamma \bar{T}] &\cong \left(2 + \frac{21}{4}\left(\frac{4}{11}\right)^{4/3}\right)\left(1 + \frac{\Omega_{b0}}{\Omega_{b0}+\Omega_{d0}} \frac{T_{eq}}{_\gamma \bar{T}}\right) & _\gamma T &\ll 1\,\mathrm{MeV} \\ g_s[_\gamma T] &\cong \frac{43}{4} & g_\rho[_\gamma T] &\cong \frac{43}{4} & 1\,\mathrm{MeV} &\ll {}_\gamma T \ll m_\mu \end{aligned} \tag{42}$$

where the 1st includes the baryon density, 2 bosonic helicity states for photons, 6 fermionic helicity states for 3 neutrino flavors (neglecting finite neutrino mass) with $_\nu \bar{T} \cong (4/11)^{4/3} \, _\gamma \bar{T}$ and the 2nd adds 4 fermionic helicity states from ultra-relativistic $e^\pm$ and has $_\nu T \cong {}_\gamma T$. Larger temperatures have larger values of $g_s$ and $g_\rho$ in the Standard Model. Matter radiation equality happens during the 1st interval and $g_s^{eq} \cong \frac{43}{11}$ while $g_\rho^{eq} \cong \left(2 + \frac{21}{4}\left(\frac{4}{11}\right)^{4/3}\right) \frac{2\Omega_{b0}+\Omega_{d0}}{\Omega_{b0}+\Omega_{d0}}$.

$\Lambda$CDM cosmological parameters from Planck 2018 (Aghanim *et al*. 2018) can be used: density parameters for matter ($\Omega_{m0} \cong 0.310$), baryons ($\Omega_{b0} \cong 0.0490$) and radiation ($\Omega_{r0} \cong 9.07 \times 10^{-5}$); the amplitude ($A_s \cong 2.10 \times 10^{-9}$) and spectral slope ($n_s \cong 0.967$) of cuvature inhomogeneities, and the current photon temperature at matter-radiation equality ($_{\gamma 0}T \cong 2.7255$ K). Derivative quantities are the dark matter density parameter $\Omega_{d0} = \Omega_{m0} - \Omega_{b0} \cong 0.261$ (Planck notation $\Omega_{c0}$), the redshift at matter-radiation equality $z_{eq} = \frac{\Omega_{m0}}{\Omega_{r0}} - 1 \cong 3413$ and the photon temperature at matter-radiation equality, $T_{eq} \cong 9304$ K $\cong 0.802$ eV, Hubble scale at matter radiation equation is $\mathcal{H}_{eq} = H_0 \sqrt{2\Omega_{m0}^2/\Omega_{r0}} = 0.01\, c/\mathrm{Mpc}$. In non-flat extended $\Lambda$CDM the empirically averaged density parameter is given by

$$\bar{\kappa} \cong \frac{(1-\Omega_0)\Omega_{r0}}{\Omega_{m0}(\Omega_{m0}+\Omega_{b0})} \frac{(1+z_{eq})^2}{(1+z)^2} = (-0.6 \pm 1.6)\times 10^{-6} \frac{(1+z_{eq})^2}{(1+z)^2} \tag{43}$$

which is empirically consistent with zero which justifies the flat $\bar{\kappa} \to 0$ assumption.

By definition when $_\gamma \bar{T} = T_{eq}$ the baryon + dark matter density equals the photon + neutrino density. If freeze-in produces a fraction $f_{fi}$ of the dark matter this determines the mean freeze-in yield

$$m_d\, \overline{Y}_f^n \cong \frac{33}{172}\left(2 + \frac{21}{4}\left(\frac{4}{11}\right)^{4/3}\right) \frac{\Omega_{d0}}{\Omega_{m0}} T_{eq} f_{fi} = 0.435\,\mathrm{eV}\, f_{fi}. \tag{44}$$

This relates the dark matter particle mass $m_d$ to the mean freeze-in yield $\overline{Y}_f^n$. Usually one would want freeze-in to produce all dark matter, $f_{fi} = 1$, but it is interesting to consider the case were only a fraction of the dark matter is produced this way. In the feeble freeze-in approxima-tion an additional pre-existing dark matter component does not invalidate the analysis given above. Note also that it has been assumed that freeze-in is completed before matter-radiation equality. This is required to be consistent with CMBR anisotropies unless $f_{\text{fi}} \ll 1$. For specif freeze-in temperature ranges these cosmological parameters give a characteristic wavenumber and mass scale

$$\mathcal{H}_x^{\text{fi}} = \begin{cases} 9.2 \text{ Mpc}^{-1} \frac{T_x^{\text{fi}}}{\text{keV}} & 1 \text{ eV} \ll T_{\text{fi}}^x \ll 1 \text{ MeV} \\ 16.3 \text{ kpc}^{-1} \frac{T_x^{\text{fi}}}{\text{MeV}} & 1 \text{ MeV} \ll T_{\text{fi}}^x \ll 100 \text{ MeV} \end{cases} \qquad M_x^{\text{fi}} = \begin{cases} 1.0 \times 10^{10} \, M_\odot \left(\frac{\text{keV}}{T_{\text{fi}}^x}\right)^3 & 1 \text{ eV} \ll T_{\text{fi}}^x \ll 1 \text{ MeV} \\ 1.8 \, M_\odot \left(\frac{\text{MeV}}{T_{\text{fi}}^x}\right)^3 & 1 \text{ MeV} \ll T_{\text{fi}}^x \ll 100 \text{ MeV} \end{cases} \tag{45}$$

where $M_x^{\text{fi}} \equiv \Omega_{d0} \rho_{\text{crit}0} (2\pi/\mathcal{H}_x^{\text{fi}})^3$ is the dark matter mass in a comoving cubical volume of sides equal to the wavelength corresponding to the wavenumber $|k| = \mathcal{H}_x^{\text{fi}}$. Implications of this characteristic quantities discussed below. As above $x = $ n, e which respectively give the dark matter and the baryon entropy anomalies. The values of characteristic quantities for $x = $ n and $x = $ e should be similar.

### IVc4 The Freeze-in Isocurvature Bump

Here we give an extension of the super-horizon result which is an *estimate* the inhomogeneous yield on scales comparable or smaller than the super-horizon scales, $|k| \ll \mathcal{H}_x^{\text{fi}}$, considered previously. To make such an estimate one can use eq 37 with a $\tilde{\kappa}[{}_s s, k]$ appropriate to these smaller scales. This is only an estimate as it treats the free-streaming neutrinos incorrectly and neglects other sub-horizon effects and in particular the free-streaming of dark matter particles which will wash out some of the inhomogeneities in the yield. Thus the following estimate of ${}_x\tilde{Y}_f[k]$ is an overestimate for $|k| \gtrsim \mathcal{H}_x^{\text{fi}}$ but not by a large factor. Approximate freeze-in is confined during an epoch when the SM matter is well approximated as an ultra-relativistic fluid such as the two cases of eq 42. Since dark matter is required long before matter-radiation equality this would be accurate except during brief periods such as during the electroweak or quark-hadron phase or when the abundance of $\mu^\pm$ or $e^\pm$ drop rapidly. For this radiation dominated case the expansion law $a[\eta] \propto \eta$ so $\mathcal{H}[\eta] = 1/\eta \propto {}_\gamma\overline{T}$. Assuming growing mode barotropic initial conditions SM inhomogeneities are acoustic waves: $\tilde{\mathcal{R}}[\eta, k] \approx 3 \tilde{\mathcal{R}}_0[k] j_1[\varphi]/\varphi$ where $\varphi = |k| \eta/\sqrt{3}$ is the acoustic phase and $j_1$ is a spherical Bessel function. Thus $\tilde{\kappa} \approx -6 \tilde{\mathcal{R}}_0 \varphi j_1[\varphi]$ and one finds

$${}_x\tilde{Y}_f[k] \approx -3 \tilde{\mathcal{R}}_0[|k|] \mathfrak{k}^2 \frac{\int_0^\infty d\varphi \, \epsilon_x\left[\frac{\mathfrak{k}}{\varphi} T_{\text{fi}}^x\right] j_1[\varphi]}{\int_0^\infty d\varphi \, \varphi \, \epsilon_x\left[\frac{\mathfrak{k}}{\varphi} T_{\text{fi}}^x\right]} \qquad \mathfrak{k} \equiv \frac{1}{\sqrt{3}} \frac{|k|}{\mathcal{H}_x^{\text{fi}}} \tag{46}$$

in terms of $\epsilon_x$ of eq 27 but expressed as a function of ${}_\gamma\overline{T}$. Since $|j_1[\varphi]| < \varphi/3$ it follows that $\left|{}_x\tilde{Y}_f\right| < \frac{1}{3} |\tilde{\mathcal{R}}_0| |k|^2 / \mathcal{H}_{\text{fi}}^{x2}$ i.e. ${}_x\tilde{Y}_f$ is smaller in magnitude than the extrapolated super-horizon formula. For $|k| \gtrsim \mathcal{H}_x^{\text{fi}}$ the numerator's integrand becomes oscillatory, greatly suppressing $\left|{}_x\tilde{Y}_f[k]\right|$ for $|k| \gg \mathcal{H}_x^{\text{fi}}$ so long as $\epsilon_x$ is a smooth function of ${}_\gamma\overline{T}$. Thus one expects the inhomogeneous yield to be peaked at $|k| \sim \mathcal{H}_x^{\text{fi}}$ with amplitude $\left|{}_x\tilde{Y}_f[k]\right| \sim |\tilde{\mathcal{R}}_0|$. The dark matter overdensity at horizon crossing for the curvature mode is $|\delta\rho/\rho|_{\text{ad}} \sim |\tilde{\mathcal{R}}_0|$ and for the freeze-in isocurvature contribution is $|\delta\rho/\rho|_{\text{iso}} \sim \left|{}_n\tilde{Y}_f\right| \sim |\tilde{\mathcal{R}}_0|$ i.e. the two are nearly the same. Thus *for all freeze-in models there is an order unity feature in the linear dark matter power spectrum* at the horizon scale during freeze-in. This feature is the *isocurvature bump*. The bump falls off at small scales (large $|k|$) because the $\kappa$ acoustic oscillations temporally average out causing the temperature-time relation to be nearly the same as the spatial average. Thus on scales much smaller than the isocurvature bump the yield is very uniform.

To illustrate this consider the class of freeze-in models where the freeze-in rates are of the form

$$\Gamma_x[{}_\gamma T] \propto {}_\gamma T^p \, e^{-T_{\text{fi}}^x/{}_\gamma T} \quad \text{or} \quad \epsilon_x[{}_\gamma T] = {}_\gamma T^{p-2} \, e^{-T_{\text{fi}}^x/{}_\gamma T} \, \epsilon_x[{}_s\overline{s}] \qquad p < 2 \tag{47}$$

which is motivated in §IV. Here $p$ regulates how rapidly DM production ramps up (smaller $p$ means faster ramp up) before it is exponentially cut-off at ${}_\gamma\overline{T} < T_{\text{fi}}^*$. For this model

$$\frac{{}_x\tilde{Y}_f[k]}{\tilde{\mathcal{R}}_0[k]} = -3 \frac{x \left(2x^2 - p(x^2-1)\right) \cos[p \tan^{-1}[x]] - (1 + (3-2p) x^2) \sin[p \tan^{-1}[x]]}{(1-p) \, x \, (1+x^2)^{\frac{4-p}{2}}} \xrightarrow{x \to \infty} -3 \frac{2-p}{1-p} \frac{\cos[p \frac{\pi}{2}]}{x^{2-p}} \tag{48}$$

where $x \equiv |k| \eta / \sqrt{3(3-p)(2-p)}$. This ratio reduces to a rational polynomial in $x$ for integer $p$. The ratio, ${}_x\tilde{Y}_f / \tilde{\mathcal{R}}_0[k]$ is plotted in Fig 1 for $p \in [-2, 1.5]$. For these models the bump peaks at $|k| = k_{\text{max}} \sim 4 \, \mathcal{H}_x^{\text{fi}}$ with peak absolute amplitude up to ${}_x\tilde{Y}_f \sim -2 \tilde{\mathcal{R}}_0$.

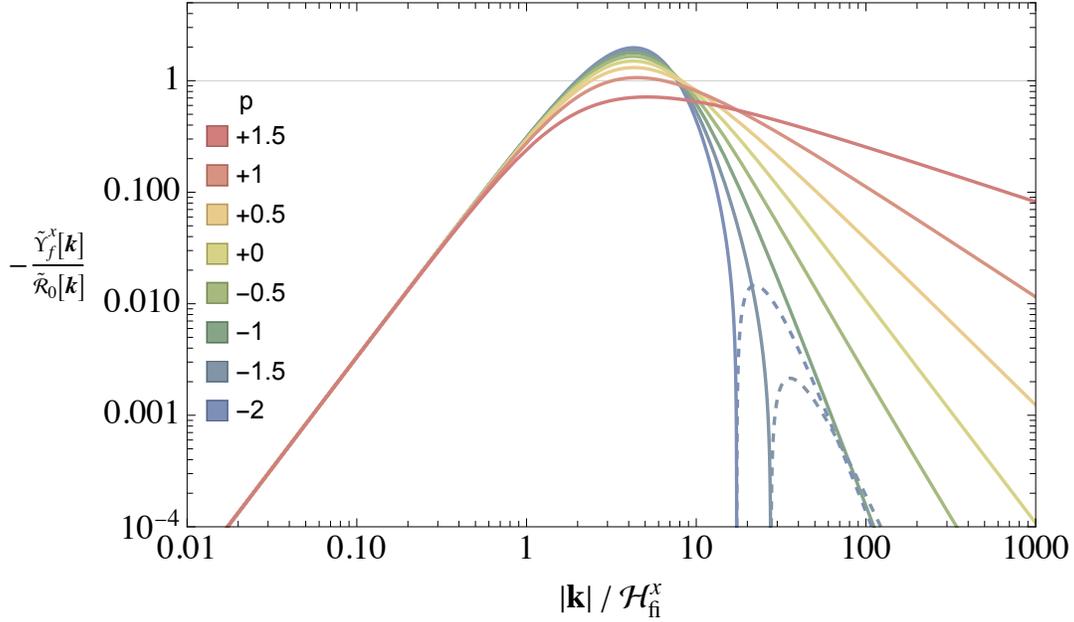

**Figure 1:** Plotted as a function of wavenumber $|k|$ are absolute value of the ratio of the Fourier amplitudes of yield inhomogeneities, $_x\tilde{Y}_f$, to primordial curvature inhomogeneities, $\tilde{\mathcal{R}}_0$, assuming the analytic freeze-in production rates of eq 47. The solid lines give negative values of this ratio and the dashed positive values. Larger negative $p$ corresponds to shorter durations of freeze-in. $\mathcal{H}_{fi}^x$ is a characteristic horizon wavenumber during freeze-in given by eq 39. Here either $x = n$ for dark matter particle production or $x = e$ for energy loss of standard matter to dark matter. $_n\tilde{Y}_f[k]$ is the Fourier amplitude of fractional dark matter to photon ratio inhomogeneities. $_e\tilde{Y}_f[k]$ is proportional to but much larger than the fractional baryon to photon ratio inhomogeneities as described in §IVc6. These curves are oscillatory if $p < -1$ with a finite number, $\max[0, \text{ceiling}[(1+p)/2]]$, of zeros.

### IVc5 Evolution of DM Inhomogeneities after Freeze-in

The yield computed previously including in Fig. 1 gives the dark matter to entropy ratio just after freeze-in but does not include any subsequent evolution which will result from relative motion of the SM and DM fluid. Such evolution must be included to compare these model with observations. Realistic modelling of this evolution would require detailed numerical radiative transfer which, while within the scope of publically available cosmological software, is outside the scope of this paper. Instead to illustrate some of the phenomenology semi-realistic but analytic models are presented. In particular, as in the previous subsection, one ignores neutrino free-streaming, treat the baryon-photon system as a perfect fluid, neglect photon viscosity, and restricts oneself to the radiation dominated epoch. The model is limited to times well after freeze-in and long before matter-radiation equality: $\mathcal{H}_{eq} \ll \eta^{-1} \ll \mathcal{H}_{fi}^n$; and to comoving wavenumbers which are are super-horizon during freeze-in and become sub-horizon during the radiation era: $\mathcal{H}_{eq} \gg |k| \gg \mathcal{H}_{fi}^n$. Here $\mathcal{H}_{eq} = \Omega_{m0}\sqrt{2/\Omega_{r0}}\,H_0/c \cong 0.0073/\text{Mpc}$ gives the comoving wavenumber of the Hubble scale at matter-radiation equality. The goal is to model these modes as they enter the horizon and oscillate acoustically up to matter radiation equality.

In this idealized freeze-in model both SM and DM are individually assumed adiabatic after freeze-in. Define the entropy density of the pressureless DM fluid as the number density of dark matter particles: $_d s = _d n$. Using $_x\tilde{\lambda} \equiv \ln[_x s /_x \bar{s}]$ the yield can be written

$$_n\overline{Y}[\eta, k] = _n\overline{Y}[\eta]\left(1 + _n\tilde{Y}[\eta, k]\right) \qquad _n\overline{Y}[\eta] = \frac{_d\bar{s}[\eta]}{_s\bar{s}[\eta]} \qquad _n\tilde{Y}[\eta, k] \cong \ln\left[1 + _n\tilde{Y}[\eta, k]\right] = \ln\left[\frac{_d s}{_s s}\right] - \ln\left[\frac{_d\bar{s}}{_s\bar{s}}\right] = _d\tilde{\lambda}[\eta, k] - _s\tilde{\lambda}[\eta, k] \quad . \tag{49}$$

The $_x\tilde{\lambda}$ are gauge dependent but $_d\tilde{\lambda} - _s\tilde{\lambda} \cong _n\tilde{Y}$ is gauge independent giving the isocurvature inhomogeneity. The final freeze-in yield provides initial conditions for the subsequent evolution: $_n\overline{Y}[\eta] \xrightarrow[fi]{} _n\overline{Y}_f$ and $_n\tilde{Y}[\eta, k] \xrightarrow[fi]{} _n\tilde{Y}_f[k]$. Since no further dark matter or entropy is produced after freeze-in the mean yield does not change after freeze-in i.e. $_n\overline{Y}[\eta] = _n\overline{Y}_f$. However the inhomogeneous dark matter to entropy ratio, $_n\tilde{Y}$, will change as the two fluids move differently since the SM fluid is accelerated by pressure gradients while the DM is not. In the feeble freeze-in scenario the DM inhomogeneities and DM peculiar velocities are both 1st order in perturbation theory and to linear order evolve independently.

For this reason one can think of the isocurvature DM inhomogeneity ($_d\tilde{\lambda} \neq {}_s\tilde{\lambda}$) as being fixed in place and superposed with an initial curvature mode which initially has $_d\tilde{\lambda} = {}_s\tilde{\lambda}$ initially and where the SM and DM are comoving. The former is the isocurvature mode and the latter the initial adiabatic barotropic SM growing mode.

After freeze-in the peculiar velocity of the SM and DM components in longitudinal gauge are

$$_s\tilde{v}[\eta, k] = i\sqrt{3}\,\tilde{\mathcal{R}}_0[k]\,\hat{k}\,\underbrace{\frac{(2+\varphi^2)j_1[\varphi] + \varphi^3 y_1[\varphi]}{1+\varphi^2}}_{\text{standard}} \qquad _d\tilde{v}[\eta, k] = i\,2\sqrt{3}\,\tilde{\mathcal{R}}_0[k]\,\hat{k}\,\underbrace{\frac{j_0[\varphi]-1}{\varphi}}_{\text{standard}} \quad (50)$$

which is independent of when or if freeze-in happened since the isocurvature mode is "at rest". Peculiar velocities are of course not locally covariant quantities however relative velocities are, e.g. $_s\tilde{v} - {}_d\tilde{v}$ which determines the evolution locally covariant isocurvature mode described by the dark matter to entropy ratio which to 1st order is

$$_d\tilde{\lambda}[\eta, k] - {}_s\tilde{\lambda}[\eta, k] \cong \underbrace{_n\tilde{Y}_f[k]}_{\text{anomalous}} + \underbrace{3\,\tilde{\mathcal{R}}_0[k]\left(\frac{\varphi^3 j_1[\varphi] + \varphi^2 y_1[\varphi]}{1+\varphi^2} + 2\,\text{Ci}[\varphi] - 2\ln[\varphi] - 2\gamma_E + 1\right)}_{\text{standard}} \quad (51)$$

where Ci is the cosine, $\gamma_E$ is Euler's constant and $_n\tilde{Y}_f \cong -\frac{1}{3}\tilde{\mathcal{R}}_0(|k|/\mathcal{H}_n^{\text{fi}})^2$ as derived above. Quantities labeled "standard" are standard radiation era acoustic modes of an initially barotropic inhomogeneities. Even standard acoustic modes develop abarotropic inhomogeneities, $_d\tilde{\lambda} \neq {}_s\tilde{\lambda}$, when the two fluids cease to be comoving $_s\tilde{v} \neq {}_d\tilde{v}$. The velocity differences is caused by pressure gradients which is usually considered a "sub-horizon" effect. The standard contribution to abarotropy starts from zero and grows. For freeze-in there is an additional "anomalous" contribution to abarotropy, $\tilde{Y}_f$, which after freeze-in remains constant and does not grow to 1st order in the radiation era. The amplitude of both the anomalous and standard contribution to $_d\tilde{\lambda} \neq {}_s\tilde{\lambda}$ are $\propto |k|^2\tilde{\mathcal{R}}_0[k]$ however the time dependence is different. The (anomalous) inhomogeneous yield dominates while the mode is sufficiently outside the sound horizon ($\varphi \ll 1$) but will become subdominant. However even the anomalous isocurvature bump abarotropy will be become sub-dominant to the standard abarotropy once these scales start oscillating inside the sound horizon ($\varphi \gtrsim 1$). This different temporal behavior and anomalous isocurvature dominance for $|k| \ll \mathcal{H}$ is another rationale for the statement that this anomalous abarotropy is a *super-horizon* isocurvature perturbation while the standard abarotropy is a *sub-horizon* isocurvature perturbation. This in spite of the fact that both have the same wavenumber scaling outside the horizon.

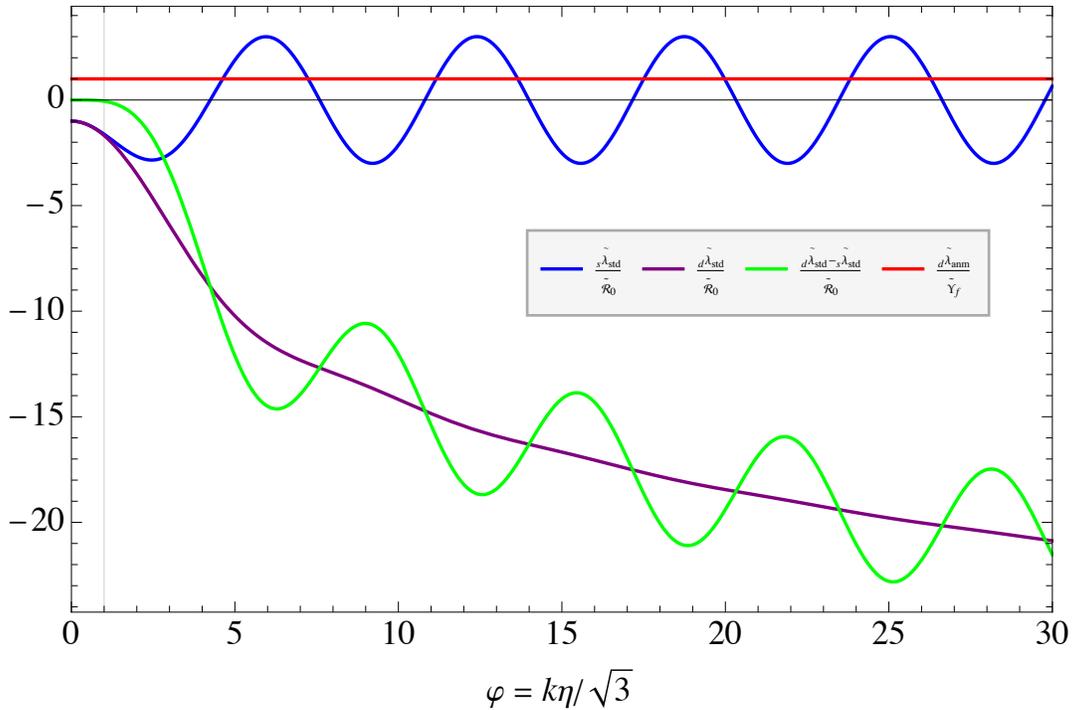

**Figure 2:** Plotted is the time evolution of fractional entropy inhomogeneities, $\tilde{\lambda} = \tilde{\delta s}/\bar{s}$ in units of the kurvature potential $\tilde{\mathcal{R}}_0$ or post freeze-in entropy ratio anomaly $\tilde{Y}_f$. $_s\tilde{\lambda}_{\text{std}}$ (blue) and $_d\tilde{\lambda}_{\text{std}}$ (purple) give the standard

values while $_d\tilde{\lambda}_{anm}$ (red) gives the contribution of the anomaly. For primordial dark matter the entropy perturbation is $\propto {_d\tilde{\lambda}_{std}} - {_s\tilde{\lambda}_{std}}$ (green) which is very small on scales above the sound horizon ($\varphi \ll 1$). For freeze-in dark matter one adds the constant $_d\tilde{\lambda}_{anm}$, i.e. the entropy perturbation is $\propto {_d\tilde{\lambda}_{anm}} + {_d\tilde{\lambda}_{std}} - {_s\tilde{\lambda}_{std}}$.

In Fig. 2 the standard and anomalous contributions to $_d\tilde{\lambda} - {_s\tilde{\lambda}}$ are plotted as the green and red curve respectively. In this graph $_d\tilde{\lambda}_{anm}$ is plotted in units of $_n\tilde{Y}_f$ while $_d\tilde{\lambda}_{std}$ or $_s\tilde{\lambda}_{std}$ are plotted in units of $\tilde{\mathcal{R}}_0$. These quantities are approximately on the same relative scale only for $|k| \sim \mathcal{H}_{fi}^n$, i.e. near the isocurvature bump. Otherwise $|_d\tilde{\lambda}_{anm}| \ll |_d\tilde{\lambda}|, |_s\tilde{\lambda}|$ at $\varphi \sim 1$. Neither $_d\tilde{\lambda}_{anm}$ not $_s\tilde{\lambda}_{std}$ grow in amplitude after sound horizon crossing, $\varphi \gg 1$, although the former remain constant and the latter oscillates. The standard isocurvature (the purple curve) grows because $_d\tilde{\lambda}_{std}$ grows logarithmically. This well known logarithmic growth is a consequence of the growing mode peculiar velocity of the DM at sound horizon crossing which decays since since DM self-gravity is negligible. Since to linear order the freeze-in isocurvature has no velocity it does not share in this growth. This logarithmic growth dilutes the isocurvature bump inhomogeneity by a logarithmic factor $\sim \ln[z_{fi}/z_{eq}]$ during radiation era.

In the matter era all DM inhomogeneities will grow by the same factor since they pick up the same velocity because they are moving in the same gravitational potential. Considering all these growth factor the freeze-in peak is diluted relative to it's amplitude just after freeze-in. At late times in the matter era but before non-linear gravitational collapse the bump has inhomogeneities enhanced by a factor $\sim 1 + C\ln[z_{fi}/z_{eq}]^{-1}$. The constant $C$ is of order unity and will depend on the specifics of freeze-in as illustrate by the model dependence of the curves in Fig. 1.

## IVd Baryon Density Isocurvature (BDI) from Freeze-in

As noted above the energy loss of SM to DM during freeze-in will enhance the baryon-to-entropy ratio and inhomogeneities in this energy loss will lead to inhomogeneities in this ratio. These inhomogeneities are baryon density isocurvature (BDI) modes. The mean baryon enhancement before and after freeze-in is given by $\ln[_b\bar{n}/_s\bar{s}]_f - \ln[_b\bar{n}/_s\bar{s}]_i = \bar{Y}_f^e$. The value of $\ln[_b\bar{n}/_s\bar{s}]_f$ has been determined with an uncertainty of $\pm 0.006$ by Planck (Aghanim *et al*. 2018) . However we have no *a priori* knowledge of the initial value and therefore no empirical way of normalizing $\bar{Y}_f^e$. This is in contrast to $\bar{Y}_f^n$ where one assumes the initial dark matter density is zero and one measures the final mass density. Nevertheless one can roughly can estimate the size of the baryon enhancements from the dark matter yield. Since the energy transfer must exceed the rest mass of dark matter particles $_s\bar{\rho}_{fi}\,_e\bar{Y}_f \gtrsim m_{d}\,_s\bar{s}_{fi}\,_n\bar{Y}_f$ where $_s\bar{\rho}_{fi}$ and $_s\bar{s}_{fi}$ are the energy and entropy density during freeze-in. Furthermore since $_s\bar{\rho}_{fi} \sim T_{fi}\,_s\bar{s}_{fi}$ one has $_e\bar{Y}_f \gtrsim (m_d/T_{fi})\,_n\bar{Y}_f$ where $T_{fi} \sim T_{fi}^n \sim T_{fi}^e$ is the temperature during freeze-in. To produce dark matter particles one requires that $T_{fi} \gtrsim m_d$ so $_e\bar{Y}_f \gtrsim (m_d/T_{fi})\,_n\bar{Y}_f \lesssim {_n\bar{Y}_f}$. Since the inequalities go in opposite directions an estimate of $_e\bar{Y}_f$ requires further assumptions. If one restricts consideration to the case where dark matter particle mass shuts off freeze in at $T_{fi} \sim m_d$ and most of the energy transfer goes to dark matter particle production then $_e\bar{Y}_f \sim {_n\bar{Y}_f} \sim (\text{eV}/m_d) f_{fi}$ from eq 44 and if $f_{fi} \sim 1$ one finds

$$\ln\left[\frac{_b\bar{n}}{_s\bar{s}}\right]_f - \ln\left[\frac{_b\bar{n}}{_s\bar{s}}\right]_i = \bar{Y}_f^e \sim \frac{\text{eV}}{m_d}. \tag{52}$$

For the mean baryon enhancement to be comparable or greater than the uncertainty in baryon density one requires $m_d \lesssim 100\,\text{eV}$.

The freeze-in induced BDI and CDI (cold dark matter density isocurvature) differ in that the fractional inhomogeneity in baryon-to-entropy ratio is $\tilde{\delta}_b^{BDI} \equiv (_bn - {_b\bar{n}})/{_b\bar{n}} \cong {_e\bar{Y}_f}\,_e\tilde{Y}_f$ whereas the fractional inhomogeneity in dark matter-to-entropy ratio is $\tilde{\delta}_d^{CDI} \equiv (_dn - {_d\bar{n}})/{_d\bar{n}} \cong {_n\tilde{Y}_f}$. The difference arises because all of the dark matter is produced by freeze-in whereas the baryons are slightly enhanced relative to the entropy by diminution of the SM density. From the estimate of $\bar{Y}_f^e$ the BDI amplitude is $\tilde{\delta}_b^{BDI} \cong {_e\bar{Y}_f}\,_e\tilde{Y}_f \sim (\text{eV}/m_d)\,_e\tilde{Y}_f$. Using the maximal isocurvature bump amplitude $_e\tilde{Y}_f \sim \tilde{\mathcal{R}}_0$ one finds that $|\tilde{\delta}_b^{BDI}| \lesssim (\text{eV}/m_d)|\tilde{\mathcal{R}}_0|$. For large $m_d \gg \text{eV}$ this would be difficult to extract from observations since it is much smaller than the curvature inhomogeneities.

## IVd Freeze-In Models and Constraints

### IVd1 Freeze-In Models

Based on formula from BBB22 approximate formulae for $\epsilon_n[T]$ and $\epsilon_e[T]$ for production of millicharged particles from $e^\pm$ annihilation are given in appendix A for millicharged particle masses in the range $\text{MeV} \ll m_d \ll m_\mu$ ($m_\chi$ is used for $m_d$ in the Appendix). These results are summarized by

$$\begin{aligned}
\overline{Y}_f^n &\approx 5.78 \times 10^{-7} \frac{\text{MeV}}{m_d} f_{fi} & Y_{eq}^n &\approx 1.34 \times 10^{-12} \left(\frac{\text{MeV}}{m_d}\right)^2 & T_{fi}^n &\approx 0.41\, m_d \\
\overline{Y}_f^e &\approx 9.43 \times 10^{-7} \frac{\text{MeV}}{m_d} f_{fi} & Y_{eq}^e &\approx 1.64 \times 10^{-12} \left(\frac{\text{MeV}}{m_d}\right)^2 & T_{fi}^e &\approx 0.37\, m_d
\end{aligned} \quad (53)$$

where $Y_{eq}^x \equiv {}_x \tilde{Y}_f[k_{eq}]$, $k_{eq} = \mathcal{H}_{eq}/c$, $f_{fi} \le 1$ is the fraction of dark matter which is frozen-in, usually 1. For $m_d \ll m_e$ millicharge production by $e^\pm$ is highly suppressed as the number of $e^\pm$ decreases by a factor $\sim 10^{-9}$ when ${}_\gamma T \ll m_e$. Thus we require $m_d \gtrsim \text{MeV}$ and $Y_{eq}^n, Y_{eq}^e \lesssim 10^{-12}$, i.e. even at the isocurvature bump the fractional isocurvature density perturbation is $\lesssim 10^{-12}$. This is many orders of magnitude below the sensitivity of the CMB and is also allowed by Ly-$\alpha$ clouds. Thus we are far from being able to constrain these freeze-in models.

**Table 1.** Temperature dependence of dark matter particle production rates and effective freeze-in temperature, $T_{fi}$, for three models of Hall et al. (2010) and for millicharged dark matter as approximated in Appendix A. $T_{fi}$ is computed using eq 41 for ultra-relativistic SM where $\epsilon_n[{}_s T] \propto {}_s T^{-5} \dot{n}[{}_s T]$. $K_1$ is a modified Bessel function.

| reaction | rate $\dot{n} \propto$ | reference | $T_{fi}^n$ |
|---|---|---|---|
| $B_1 \to B_2 + d$ | ${}_s T\, K_1\!\left[\frac{m_{B_1}}{{}_s T}\right]$ | Hall *et al.* eq 6.6 | $\frac{m_{B_1}}{\sqrt{15}}$ |
| $B_1 + B_2 \to d$ | ${}_s T\, K_1\!\left[\frac{m_d}{{}_s T}\right]$ | Hall *et al.* eqs 6.6 & 6.16 | $\frac{m_d}{\sqrt{15}}$ |
| $B_1 + B_2 \to B_3 + d$ | ${}_s T^3\, K_1\!\left[\frac{m_d}{{}_s T}\right]$ | Hall *et al.* eq 6.24 | $\frac{m_d}{\sqrt{3}}$ |
| $e^+ + e^- \to d + \bar{d}$ | ${}_s T^4 \left( e^{-\frac{1}{2}\frac{m_d}{{}_s T}} + \frac{3}{2}\left(\frac{\pi m_d}{{}_s T}\right)^{1/2} e^{-2\frac{m_d}{{}_s T}} \right)$ | eq A4 | $\frac{m_d}{2^{3/2}} \sqrt{\frac{1+\frac{3\pi}{2^{9/2}}}{1+\frac{45\pi}{2^{23/2}}}}$ |

There are many other dark matter particle candidates which could be produced by the freeze-in mechanism. This mechanism was first proposed in Hall *et al.* (2010) where freeze-in was calculated for three models. Table 1 lists the particle production rates for these three models as well as the millicharge model and lists the corresponding freeze-in temperature $T_{fi}^n$. In the 1st model $T_{fi}^n \propto m_{B_1}$ where $m_{B_1}$ is the mass of $B_1$ which is an intermediate particle and it is possible that $m_{B_1} \gg m_\chi$ so one doesn't have the mass scaling of eq 53. However in the last 3 models the $m_\chi$ scaling is satisfied in which case

$$r \equiv \frac{T_{fi}^n}{m_\chi} \qquad Y_{eq}^n \cong 0.5 \frac{g_s^{fi\,2/3}}{g_\rho^{fi}} \left(\frac{\text{eV}}{r m_d}\right)^2 \cong \begin{cases} 3.70 \times 10^{-7} \left(\frac{\text{keV}}{r m_d}\right)^2 & \text{eV} \ll T_{fi}^n \ll \text{MeV} \\ 2.26 \times 10^{-13} \left(\frac{\text{MeV}}{r m_d}\right)^2 & \text{MeV} \ll T_{fi}^n \ll m_\mu \end{cases}. \quad (54)$$

For the reaction rates of Table 1 the factor $r^{-2}$ varies between 3 and 15 which may be typical of freeze-in models with $T_{fi}^n \propto m_d$. For freeze modes limited by the mass of an intermediary particle masses $r^{-2}$ will likely be smaller since the intermediary is likely more massive than the dark matter particle. Thus $Y_{eq}^n$ is likely a small number if $m_\chi \gg 1$ eV. For MeV $\ll T_{fi}^n$

**IVd1 Constraints on Freeze-in**

Viable dark matter candidates must be cold not warm, *i.e.* have sufficiently low velocity dispersion: roughly that they are very non-relativistic at $z \sim z_{eq}$. In freeze-in scenarios, since the dark matter particles never attain thermal equilibrium, the velocity distribution is not thermal and must be calculated and compared with observational data to determine whether or not it is viable. The most stringent limits come from the existence of low mass Ly-$\alpha$ clouds whose production would be suppressed if the dark matter is too warm. The translation of a dark matter velocity distribution to Ly-$\alpha$ cloud observables is itself non-trivial. Such a study has been carried out by d'Eramo and Lenoci (2021). As with the isocurvature modes discussed here the relation between the dark matter mass $m_d$ and the constrained quantities depends on details of the DM-SM interactions. However roughly speaking $T_{fi}^x$, $m_d \lesssim 10$ keV is excluded by Ly-$\alpha$ data.

The constraint on the epoch of freeze-in pushes the wavenumber of the isocurvature bump to $\mathcal{H}_{fi}^x \gtrsim 1/\text{Mpc}$ and the factor giving the logarithmic dilution of the bump to $\ln[z_{fi}/z_{eq}] \gtrsim 10$. Thus the isocurvature bump is at most a $\sim$10% enhancement in amplitude (20% in power) on a scale which is today highly non-linear. The Mpc scale is roughly the mass scale of small galaxies and is easily accessible observationally. The difficulty in detecting isocurvature bump feature is non-linearity since extrapolating objects of this mass scale back in time to when they first collapsed with 10% accuracy is challenging; but perhaps not impossible.

While the isocurvature bump is constrained to occur at scales smaller than resolved by CMB observations the anomalous isocurvature modes extend to arbitrarily large (super-horizon) scales where non-linearity is not an issue. The amplitude of anomalous isocurvature modes on these larger scales is much smaller than at the bump but precise CMB observations are more sensitive to small modifications of the standard model ($\Lambda$CDM) curvature inhomogeneities. A detailed comparison with CMB data is beyond the scope of this paper and left to other work. From eq 54 we see the velocity dispersion constraint requires $\Upsilon^n_{eq} \lesssim 10^{-8}$. It is not expected that current data would be sensitive enough to detect this very small contribution to cosmic inhomogeneities. Given the larger amplitude at isocurvature bump scales and the linearity of the CMB suggests that freeze-in of dark matter at the $\sim 10$ keV scale might be accessible to CMB measurements with better angular resolution than is currently used.

# IV. Synopsis

In this paper it is shown how curvature inhomogeneities can deterministically produce isocurvature inhomogeneities on super-horizon scales when chemical processes (processes with convert one type of matter into another) do not proceed to thermal equilibrium. It has proven useful to approach this subject by restricting the analysis to local covariant quantities (a more restrictive class than "gauge invariant" quantities). With local covariant quantities it is straightforward to address issues of causality. The covariant approach generally uses non-perturbative quantities and most directly computes $n$-point correlation functions rather than spatial power spectra (see §III). Of central importance is, $K \equiv 8 \pi G \rho/3 - \theta^2/9$, called "kurvature" here, which is a non-perturbative local covariant scalar associated with the more commonly used "curvature" $\mathcal{R}$ (see §IIb2). Non-zero correlations of $K$ at causally disconnected events seed isocurvature inhomogeneities at these same events. Consider the final ratio of the number density of dark matter particles to the standard model entropy density. This is a local covariant quantity which would be uniform in a scenario with only curvature inhomogeneities. However for dark matter production after inflation the correlated curvature inhomogeneities will cause this ratio to have non-zero covariance in fluid elements which have not been in causal contact since the end of inflation. This does not violate causality because these two fluid elements were in causal contact during inflation.

This is phenomena is illustrated with the example of dark matter freeze-in for two isocurvature modes: the ratio of dark matter to SM entropy/photons and the ratio of baryons to SM entropy/photons. Analysis was made using two methodologies: separate universe modelling and cosmological perturbation theory. While the acausal isocurvature inhomogeneities do exist in these models their amplitude is very small. At a given scale / wavenumber the amplitude of the isocurvature modes become increasingly small as the redshift of freeze-in increases. The latest time at which freeze-in could occur is limited by the required smallness of the dark matter velocity dispersion. This generally limits the freeze-in temperature to $\gtrsim 10$ keV and freeze-in redshift $z \gtrsim 10^7$. This lower limit sets an upper bound on the isocurvature amplitudes which are so small that it would be challenging to measure. The amplitude of the isocurvature mode relative to the curvature mode is maximized as the horizon scale during freeze-in which we call the "isocurvature bump". The enhancement of dark matter inhomogeneities at the isocurvature bump is of order unity just after freeze-in but is logarithmically diluted while inside the horizon to $\lesssim 10\%$. The previously mentioned velocity dispersion constraints pushes the isocurvature bump to non-linear scales $\lesssim 1$ Mpc where it would be difficult to disentangle a feature in the power spectrum from uncertainties in baryonic physics. Precise limits from small scale clustering as well as from large scale CMB are beyond the scope of this paper but it is expected that current observations of small scale clustering are not competitive with Ly-$\alpha$ cloud in constraining dark matter freeze-in.

§II of this paper gives a pedagogic commentary on some aspects of cosmological perturbation theory, including the local covariant approach. §IIc and §IId contains a discussion of the semantics of certain words by cosmologists. The usage of the words "adiabatic" and "clock" are called out for criticism.

Note added after completion: Some aspects of this work are mirrored in the independent work of Holst Hu & Jenks (2023).

# Acknowledgements

The author thanks Kim Berghaus, Davide Racco, Tony Riotto, Tanner Trickle, Rodolfo Capdevilla and Gordan Krnjaic for useful discussion on various aspects of this paper. Special thanks to Berghaus for getting me interested in this topic. Thus work was fully supported by the Fermi Research Alliance, LLC under Contract No. DE-AC02-07CH11359 with the U.S. Department of Energy, Office of Science, Office of High Energy Physics.

# Appendix A. MeV Millicharge Freeze-In

Here is given approximations to millicharged particle production using reaction rate (BBB22) for $e^+ + e^- \rightarrow \chi + \bar{\chi}$ production appropriate for $m_e \ll m_\chi \ll m_\mu$:

$$\chi\dot{n} = 2 \times 4\pi \int_0^\infty dp\, \frac{p^2}{\sqrt{p^2 + m_\chi^2}}\, C[p]. \qquad \chi\dot{e} = 2 \times 4\pi \int_0^\infty dp\, p^2\, C[p] \qquad (1)$$

$$C[p] = \frac{Q_\chi^2 e^4 T}{24 \pi^3 p} \int_{s_{\min}}^{\infty} ds\, e^{-\frac{s\sqrt{p^2+m_\chi^2}}{2 m_\chi^2 T}} \sinh\left[\frac{p\sqrt{s(s-4m_\chi^2)}}{2 m_\chi^2 T}\right] \sqrt{1 - \frac{4 m_\chi^2}{s}} \left(1 + \frac{2 m_e^2}{s}\right)\left(1 + \frac{2 m_\chi^2}{s}\right) \quad (2)$$

where $T$ is the SM plasma temperature, $m_\chi$ is the millicharged particle mass, $Q_\chi e$ is the electrical millicharge, $p$ particle momentum, $\sqrt{s}$ is the center of mass energy and $s_{\min} = \max\left[4 m_\chi^2, 2 m_\chi \sqrt{p^2 + m_\chi^2}\right]$. The ultra-relativistic $(m_e \ll m_\chi \ll T \ll m_\mu)$ and non-relativistic $(m_e \ll T \ll m_\chi \ll m_\mu)$ limits are

$$C[p] \cong \frac{Q_\chi^2 e^4 m_\chi^2}{48 \pi^3} \begin{cases} \frac{T}{p} \int_{2 m_\chi}^{\infty} \frac{ds}{p\, m_\chi^2}\, e^{-\frac{p}{T}} e^{-\frac{s}{4pT}} = 4 \frac{T^2}{m_\chi^2} e^{-\frac{m_\chi}{2T}} e^{-\frac{p}{T}} & T \gg m_\chi \\ 24 \int_0^\infty d\Delta\, \Delta\, e^{-2\frac{\sqrt{p^2+m_\chi^2}}{T}(1+\Delta)} \cong 6 \frac{T^2}{m_\chi^2} e^{-2\frac{m_\chi}{T}} e^{-\frac{p^2}{m_\chi T}} & T \ll m_\chi \end{cases} \quad (3)$$

so

$$\begin{aligned}
_d\dot{\tilde{n}} &\approx \frac{Q_\chi^2 e^4 T^4}{6\pi^2} \begin{cases} e^{-\frac{1}{2}\frac{m_\chi}{T}} & T \gg m_\chi \\ \frac{3}{2}\left(\frac{\pi m_\chi}{T}\right)^{1/2} e^{-2\frac{m_\chi}{T}} & T \ll m_\chi \end{cases} \\
_d\dot{\tilde{e}} &= \frac{Q_\chi^2 e^4 T^5}{3\pi^2} \begin{cases} e^{-\frac{1}{2}\frac{m_\chi}{T}} & T \gg m_\chi \\ \frac{3}{4\pi}\left(\frac{\pi m_\chi}{T}\right)^{3/2} e^{-2\frac{m_\chi}{T}} & T \ll m_\chi \end{cases}
\end{aligned} \quad (4)$$

A reasonable interpolation between these two asymptotic forms is simply to sum the two since each term dominates in its regimes of applicability. This interpolation using $g_s^{fi} \cong g_\rho^{fi} \cong \frac{43}{4}$ for $m_e \ll T \ll m_\mu$ is

$$\begin{aligned}
\epsilon_n[T] &\approx \frac{45}{43} \sqrt{\frac{5}{43\pi^{11}}}\, \frac{e^4 Q_\chi^2}{\sqrt{G}\, T} \left(e^{-\frac{1}{2}\frac{m_\chi}{T}} + \frac{3}{2}\left(\frac{\pi m_\chi}{T}\right)^{1/2} e^{-2\frac{m_\chi}{T}}\right) \\
\epsilon_e[T] &\approx \frac{90}{43} \sqrt{\frac{5}{43\pi^{11}}}\, \frac{e^4 Q_\chi^2}{\sqrt{G}\, T} \left(e^{-\frac{1}{2}\frac{m_\chi}{T}} + \frac{3}{4\pi}\left(\frac{\pi m_\chi}{T}\right)^{3/2} e^{-2\frac{m_\chi}{T}}\right)
\end{aligned} \quad (5)$$

so

$$\begin{aligned}
\overline{Y}_f^n &\approx \frac{90}{43} \sqrt{\frac{5}{43\pi^{11}}} \left(1 + \frac{3\pi}{2^{9/2}}\right) \frac{e^4 Q_\chi^2}{\sqrt{G}\, m_\chi} & T_{fi}^n &\approx \frac{m_\chi}{2^{3/2}} \sqrt{\frac{1 + \frac{3\pi}{2^{9/2}}}{1 + \frac{45\pi}{2^{23/2}}}} \\
\overline{Y}_{fb}^e &\approx \frac{180}{43} \sqrt{\frac{5}{43\pi^{11}}} \left(1 + \frac{9\pi}{2^{15/2}}\right) \frac{e^4 Q_\chi^2}{\sqrt{G}\, m_\chi} & T_{fi}^e &\approx \frac{m_\chi}{2^{3/2}} \sqrt{\frac{1 + \frac{9\pi}{2^{15/2}}}{1 + \frac{315\pi}{2^{29/2}}}}
\end{aligned} \quad (6)$$

so $\overline{\Delta\eta}_b \cong 2\,\overline{Y}$. The $T \ll m_\chi$ production terms from eq A4 contributes the most to these quantities.

Using $e^2 = \frac{\alpha_e}{4\pi}$ and the dark matter normalization of §IVc3 one requires

$$Q_\chi \approx \frac{1}{\alpha_e} \sqrt{\frac{176}{45}} \sqrt{\frac{43}{5}} \pi^{15}\, \frac{1 + \frac{21}{8}\left(\frac{4}{11}\right)^{4/3}}{1 + \frac{3\pi}{2^{9/2}}}\, \frac{\Omega_{d0}}{\Omega_{r0}}\, \sqrt[4]{G\, T_{\gamma 0}^2} \cong 2.74 \times 10^{-10} \quad (7)$$

so

$$\begin{aligned}
\epsilon_n[T] &\approx \frac{33}{172}\, \frac{1 + \frac{21}{8}\left(\frac{4}{11}\right)^{4/3}}{1 + \frac{3\pi}{2^{9/2}}}\, \frac{\Omega_{d0}}{\Omega_{m0}}\, \frac{T_{eq}}{T} \left(e^{-\frac{1}{2}\frac{m_\chi}{T}} + \frac{3}{2}\left(\frac{\pi m_\chi}{T}\right)^{1/2} e^{-2\frac{m_\chi}{T}}\right) \\
\epsilon_e[T] &\approx \frac{33}{86}\, \frac{1 + \frac{21}{8}\left(\frac{4}{11}\right)^{4/3}}{1 + \frac{3\pi}{2^{9/2}}}\, \frac{\Omega_{d0}}{\Omega_{m0}}\, \frac{T_{eq}}{T} \left(e^{-\frac{1}{2}\frac{m_\chi}{T}} + \frac{3}{4\pi}\left(\frac{\pi m_\chi}{T}\right)^{3/2} e^{-2\frac{m_\chi}{T}}\right)
\end{aligned} \quad (8)$$